\documentclass{emulateapj}
\usepackage{amsfonts}

\newcommand{\eg}{\emph{e.g.}}
\newcommand{\kms}{km\,s$^{-1}$}

\begin{document}

\submitted{Accepted for publication in ApJ, January 2017}

\title{Thermal physics of the inner coma: ALMA studies of the methanol distribution and excitation in comet C/2012 K1 (PanSTARRS)}

\author{M. A. Cordiner\altaffilmark{1,2}, N. Biver\altaffilmark{3}, J. Crovisier\altaffilmark{3}, D. Bockel{\'e}e-Morvan\altaffilmark{3}, M. J. Mumma\altaffilmark{1}, S. B. Charnley\altaffilmark{1}, G. Villanueva\altaffilmark{1}, L. Paganini\altaffilmark{1,2}, D. C. Lis\altaffilmark{4}, S. N. Milam\altaffilmark{1}, A. J. Remijan\altaffilmark{5}, I. M. Coulson\altaffilmark{6}, Y.-J. Kuan\altaffilmark{7,8}, J. Boissier\altaffilmark{9}}


\altaffiltext{1}{Goddard Center for Astrobiology, NASA Goddard Space Flight Center, 8800 Greenbelt Road, Greenbelt, MD 20771, USA.}
\email{martin.cordiner@nasa.gov}
\altaffiltext{2}{Department of Physics, Catholic University of America, Washington, DC 20064, USA.}
\altaffiltext{3}{LEISA, Observatoire de Paris, CNRS, UPMC, Universit{\'e} Paris-Diderot, 5 place Jules Janssen, 92195 Meudon, France.}
\altaffiltext{4}{LERMA, Observatoire de Paris, PSL Research University, CNRS, Sorbonne Universit{\'e}s, UPMC Univ. Paris 06, F-75014, Paris, France.}
\altaffiltext{5}{National Radio Astronomy Observatory, Charlottesville, VA 22903, USA.}
\altaffiltext{6}{Joint Astronomy Centre, Hilo, HI 96720, USA.}
\altaffiltext{7}{National Taiwan Normal University, Taipei 116, Taiwan, ROC.}
\altaffiltext{8}{Institute of Astronomy and Astrophysics, Academia Sinica, Taipei 106, Taiwan, ROC.}
\altaffiltext{9}{IRAM, 300 Rue de la Piscine, 38406 Saint Martin d'Heres, France.}

\begin{abstract}

We present spatially and spectrally-resolved observations of CH$_3$OH emission from comet C/2012 K1 (PanSTARRS) using The Atacama Large Millimeter/submillimeter Array (ALMA) on 2014 June 28-29.  Two-dimensional maps of the line-of-sight average rotational temperature ($T_{rot}$) were derived, covering spatial scales $0.3''-1.8''$ (corresponding to sky-projected distances $\rho\sim500$-2500~km). The CH$_3$OH column density distributions are consistent with isotropic, uniform outflow from the nucleus, with no evidence for extended sources of CH$_3$OH in the coma.  The $T_{rot}(\rho)$ radial profiles show a significant drop within a few thousand kilometers of the nucleus, falling from about 60~K to 20~K between $\rho=0$ and 2500~km on June 28, whereas on June 29, $T_{rot}$ fell from about 120~K to 40~K between $\rho=$~0~km and 1000~km.  The observed $T_{rot}$ behavior is interpreted primarily as a result of variations in the coma kinetic temperature due to adiabatic cooling of the outflowing gas, as well as radiative cooling of the CH$_3$OH rotational levels. Our excitation model shows that radiative cooling is more important for the $J=7-6$ transitions (at 338~GHz) than for the $K=3-2$ transitions (at 252~GHz), resulting in a strongly sub-thermal distribution of levels in the $J=7-6$ band at $\rho\gtrsim1000$~km. For both bands, the observed temperature drop with distance is less steep than predicted by standard coma theoretical models, which suggests the presence of a significant source of heating in addition to the photolytic heat sources usually considered.

\end{abstract}


\section{Introduction} \label{sec:intro}

Comets are considered fossils of the early Solar System --- frozen relics containing ice, dust and debris from the protoplanetary accretion disk. Having existed in a relatively quiescent state since their formation \citep[\eg][]{dav16}, cometary compositions can provide unique information on the thermal and chemical characteristics of the early Solar System. Most of our knowledge on cometary compositions comes from remote (ground-based) observations of their gaseous atmospheres/comae \citep{coc15}, for which the (typically relatively low) angular resolution and incomplete spatial coverage limits the amount of information that can be obtained. A particular problem is the lack of understanding regarding the physical and chemical structure of the near-nucleus coma, at distances less than a few thousand kilometres from the comet's surface.

With the advent of the Atacama Large Millimeter/submillimeter Array, high-sensitivity, high angular-resolution millimeter-wave interferometry of typical, moderately bright comets has become possible.  ALMA's unique capabilities allow us to probe the physical and chemical structure of the innermost regions of the coma in unprecedented detail, leading to new insights into the properties of the coma and the nucleus.  The first cometary observations using ALMA were reported by \citet{cor14}, who measured the distributions of HCN, HNC and H$_2$CO in comets C/2012 F6 (Lemmon) and C/2012 S1 (ISON) and demonstrated unequivocally that HNC and H$_2$CO are released in the coma (as a result of photolytic and/or thermal processes), whereas HCN originates from (or very near to) the nucleus. 

For simplicity, a constant coma kinetic temperature is commonly assumed during analysis of microwave and sub-mm cometary observations. However, theoretical and observational studies increasingly show that temperatures can vary substantially over short distance scales in the coma. Strong variations in temperature within a few hundred kilometers of the nucleus are predicted by coma hydrodynamic/Monte Carlo simulations \citep[\eg][]{kor87,com99b,com99,rod02,ten08}, but these models are largely untested due to a lack of comparative observational data. Early observational reports of inner coma temperature variations were based on long-slit infrared spectroscopy of HCN and CO in the unusually active comet C/1995 O1 (Hale-Bopp) \citep{mag99,dis01}. In a far-IR study of H$_2$O and HDO emission from comet C/2009 P1 (Garradd) by \citet{boc12}, a strongly variable temperature law was suggested --- reaching a minimum at (4-20)$\times10^3$~km from the nucleus and rising to 150~K in the outer coma. Advances in ground-based infrared instrumentation and data analysis techniques permitted \citet{bon07,bon08,bon13,bon14} to detect significant variations in the H$_2$O rotational excitation temperature over distance scales $\sim20$-1500~km in four comets, which led to the first detailed (quantitative) comparisons with theory for comet 73P/Schwassmann-Wachmann 3 \citep{fou12}. 

To-date, a general lack of information concerning spatial variations in the excitation of cometary molecules impedes the accuracy of important results in the cometary literature, hindering the derivation of accurate cometary mixing ratios, as well as introducing errors into the parent scale lengths of distributed sources. Thus, detailed measurements of coma temperatures are required in order to confirm and expand upon the previous observational findings, to stimulate revision and refinement of theoretical models.

By virtue of its large abundance in comets (typically on the order of a few percent with respect to H$_2$O), and its strong rotational bands throughout the mm/sub-mm, methanol (CH$_3$OH), is an ideal molecule for mapping the coma temperature distribution. Here we present results exploting the high resolution and sensitivity of ALMA to provide new information on the distribution and excitation of CH$_3$OH in the inner coma of the Oort-cloud comet C/2012 K1 (PanSTARRS), {and interpret our observations using non-LTE radiative transfer models}.

\section{ALMA CH$_3$OH observations}

ALMA observations of C/2012 K1 (PanSTARRS) were obtained pre-perihelion during 2014-06-28 19:07-20:05 and 2014-06-29 17:37-18:26, while the comet was at a heliocentric distance $r_H=1.42$-1.43~AU and geocentric distance $\Delta=1.96$-1.97~AU (the comet reached perihelion at $r_H=1.05$~AU on 2014-08-28). The CH$_3$OH $K=3-2$ rotational band near 251.9~GHz (in ALMA receiver band 6) was observed on June 28 and the $J=7-6$ band near 338.5~GHz (in band 7) was observed on June 29. Atmospheric conditions were excellent throughout (with zenith PWV~$<0.4$~mm). Thirty 12-m antennae, with baseline lengths 20-650~m,  resulted in an angular resolution of $0.80''\times0.43''$ at 252~GHz and $0.71''\times0.33''$ at 338~GHz. These beam dimensions correspond to $1100\times610$~km and $1000\times470$~km, respectively, at the distance of the comet and the measured RMS noise levels per channel were 1.8~mJy\,bm$^{-1}$ and 3.0~mJy\,bm$^{-1}$. The correlator was configured to simultaneously observe as many strong CH$_3$OH lines as possible (spanning different upper-state energy levels) in a single 976 kHz spectral window, using a spectral resolution of 488~kHz {(0.58~\kms\ in band 6 and 0.43~\kms\ in band 7)}. 

The observing sequence on each date consisted of an interleaved series of scans of the science target and a continuum source (quasar) for phase calibration, alternating between 7 minutes integration on the comet and 30~s on the phase calibrator, for a total of 46 min on-source at 252~GHz and 38 min on-source at 338~GHz. The comet was tracked, and the position of the array phase center updated in real-time using the latest JPL Horizons orbital solution (sampled at 15~s intervals and interpolated using a 4th-order polynomial). The data were flagged, calibrated and cleaned using standard CASA routines \citep[see for example][]{cor14}, with Ganymede as the flux calibrator. Imaging was performed using a grid size of $768\times768$ pixels, with $0.05''$ pixel scale. The resulting data cubes were corrected for the response of the ALMA primary beam and then transformed from celestial coordinates to sky-projected distances with respect to the center of the comet. Spectral fluxes per beam were subsequently converted to the Rayleigh-Jeans brightness temperature scale ($T_B$) for further analysis.

\section{Nan\c{c}ay OH observations}
\label{sec:oh}

\begin{figure}
\centering
\includegraphics[width=\columnwidth]{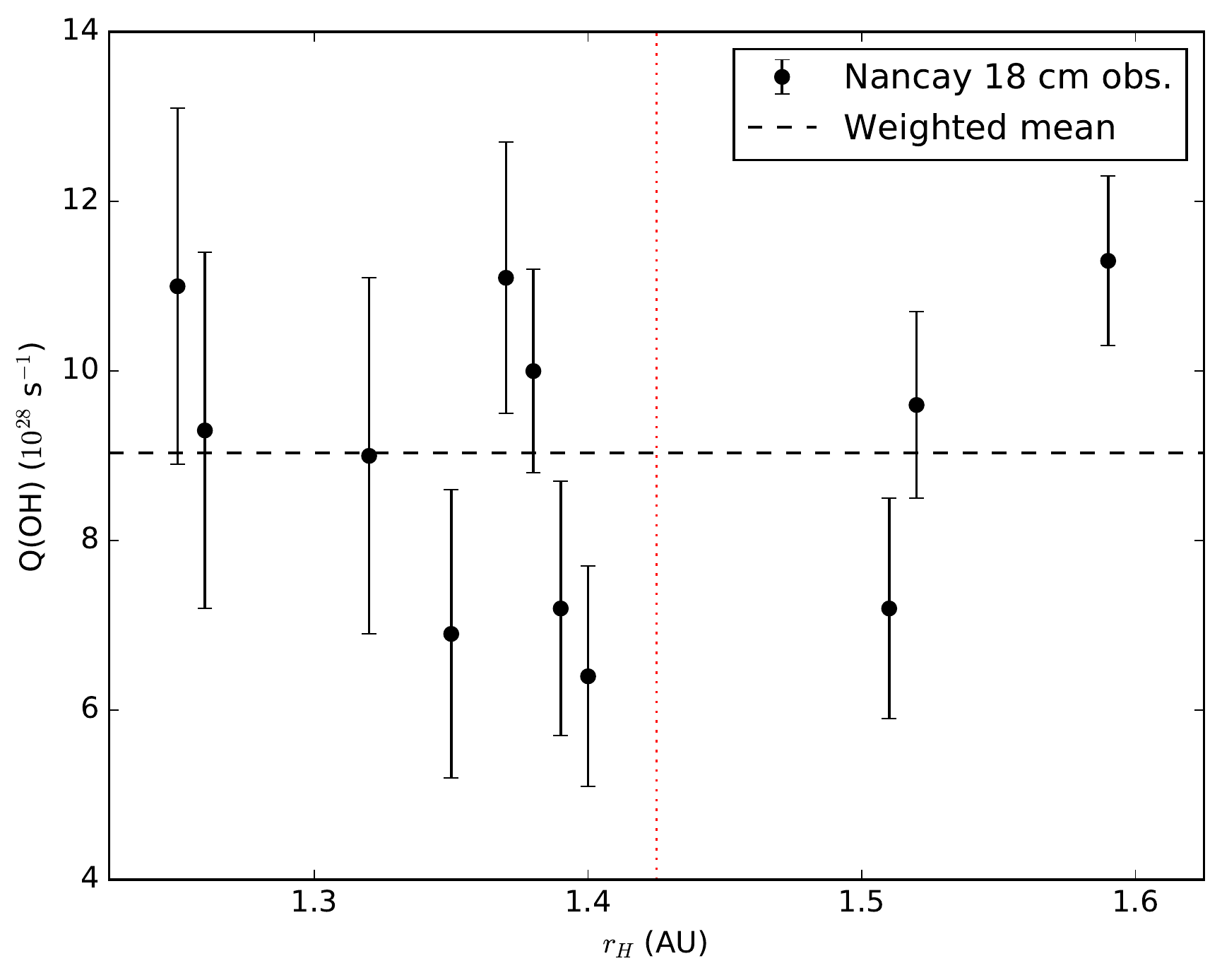}
\caption{Production rates for OH in comet C/2012 K1 (PanSTARRS) as a function of heliocentric distance obtained using the Nan\c{c}ay radio telescope. {The values shown are pre-perihelion, and cover the period 2014-06-14 to 2014-07-17}. Dashed horizontal line shows the mean value of $9.0\times10^{28}$~s$^{-1}$. Vertical dotted line corresponds to the epoch of the ALMA observations. \label{fig:oh}}
\end{figure}

Production rates for the dominant volatile H$_2$O are required for the measurement of CH$_3$OH mixing ratios and the derivation of the collisional excitation rates. The strengths of the 18~cm OH lines (at 1667 and 1665 MHz) in comet C/2012 K1 (PanSTARRS) were monitored using the Nan\c{c}ay radio telescope during the period 2014-04-03 to 2014-09-27. The observational procedure and data analysis were described previously by \citet{cro02}, and an expansion velocity of 0.70~\kms\ was retrieved following the analysis method of \citet{tse07}.  The resulting OH production rates ($Q({\rm OH})$, for the period 2014-06-14 to 2014-07-17) are shown as a function of heliocentric distance (pre-perihelion) in Figure \ref{fig:oh}, including error bars due to thermal noise. The OH line inversion parameter was large for these observations, therefore errors on the OH production rate due to excitation uncertainties are expected to be negligible. There is no clear trend in $Q({\rm OH})$ \emph{vs.} $r_H$ {around the time of our ALMA observations}, so we take the mean value of $9.0\times10^{28}$~s$^{-1}$ {(with a standard deviation of $1.7\times10^{28}$~s$^{-1}$). Assuming H$_2$O is the sole parent of OH, and adopting a branching ratio of 0.86 for the H$_2$O +~$h\nu$ $\longrightarrow$~OH +~H photolysis channel \citep{hue92}}, we obtain a water production rate of $(1.05\pm0.20)\times10^{29}$~s$^{-1}$.

\section{Results}

\subsection{Spectra and maps}

\begin{figure*}
\centering
\includegraphics[width=0.45\textwidth]{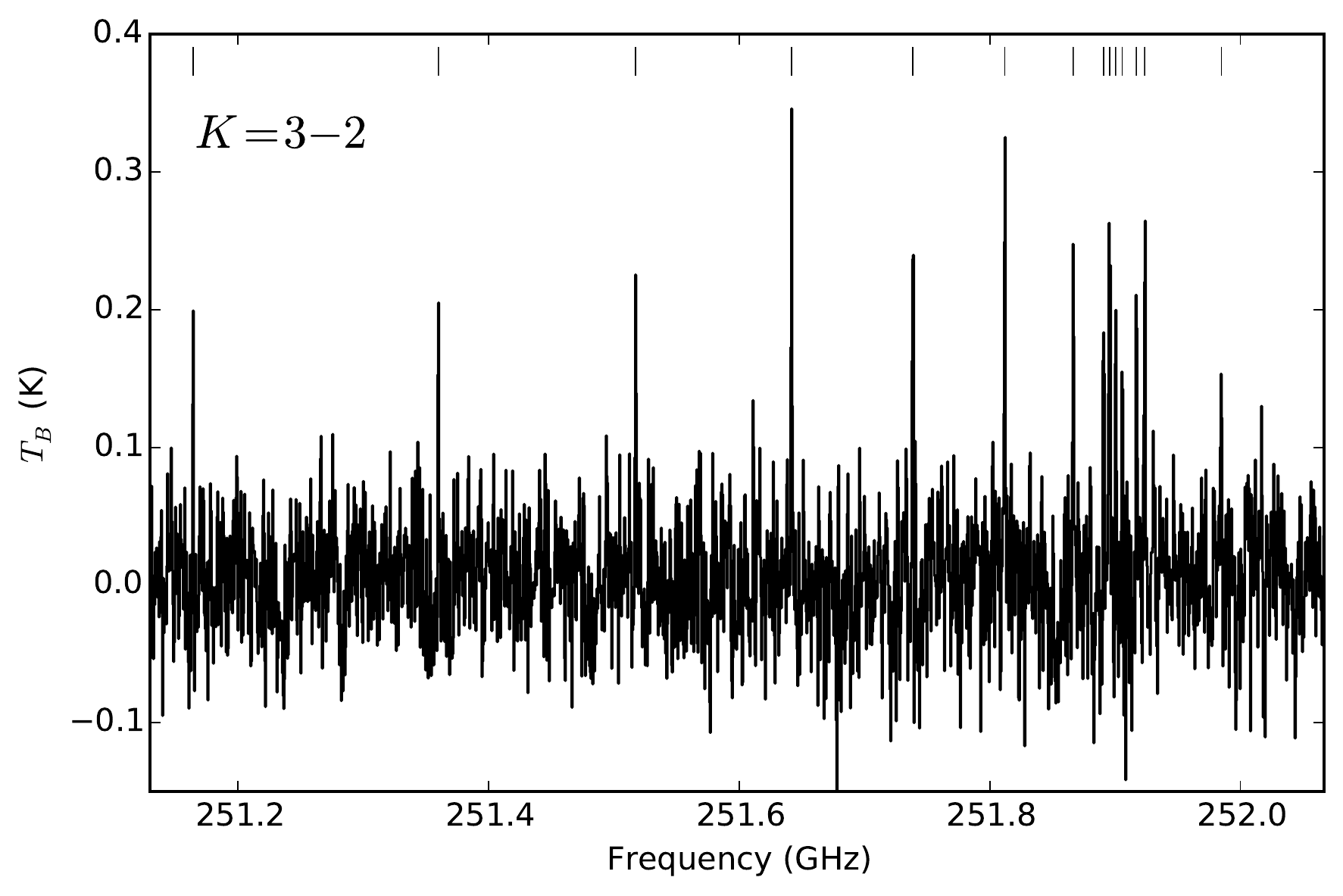}
\hspace{4mm}
\includegraphics[width=0.45\textwidth]{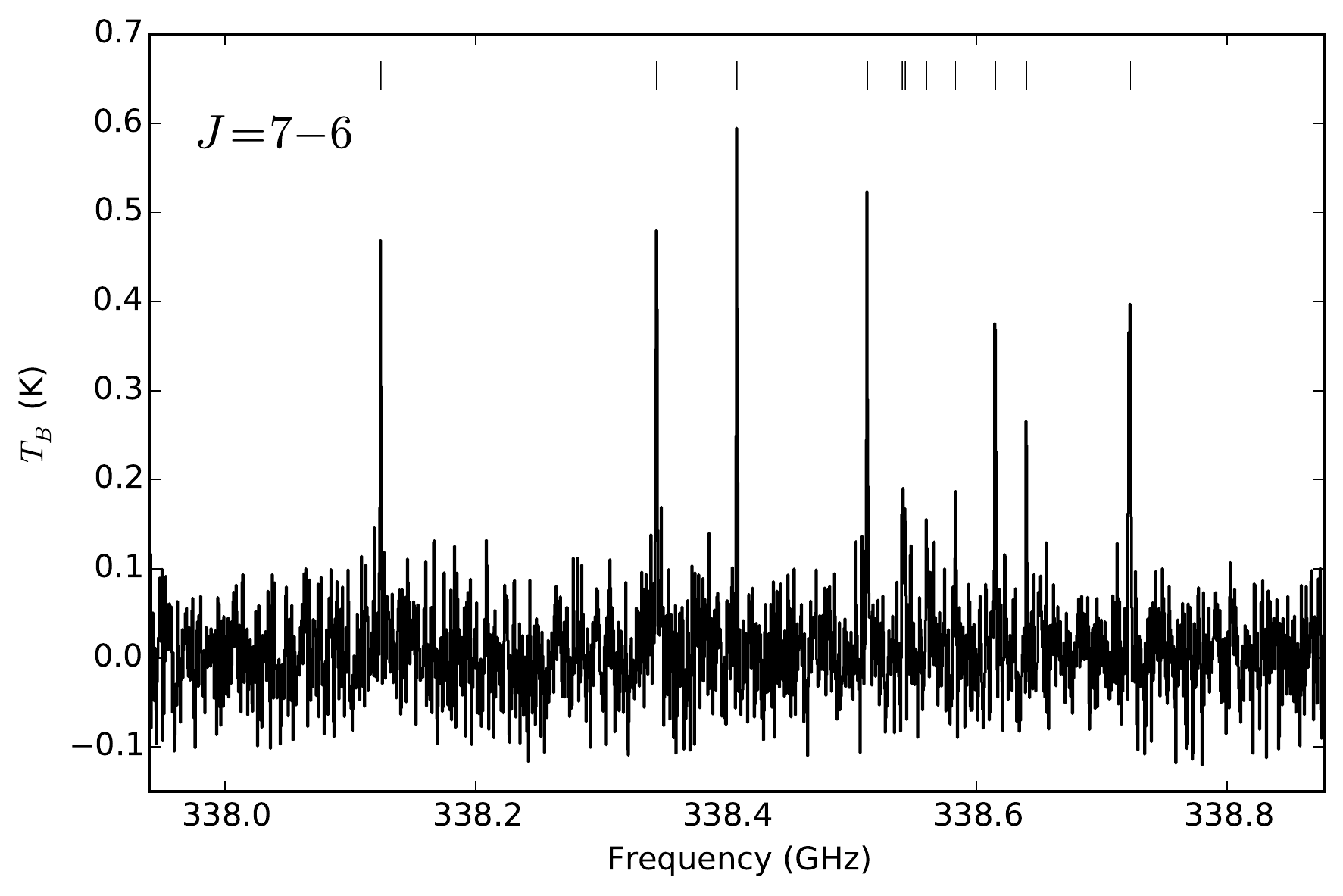}
\caption{CH$_3$OH spectra observed on 2014-06-28 of the $K=3-2$ band (left) and on 2014-06-29 of the $J=7-6$ band (right). Ticks indicate the detected CH$_3$OH lines. These spectra were averaged over a $1.5''$-diameter circle ($\approx2100$~km) centered on the emission peak.\label{fig:spec}}
\end{figure*}

\begin{figure*}
\centering
\includegraphics[width=0.45\textwidth]{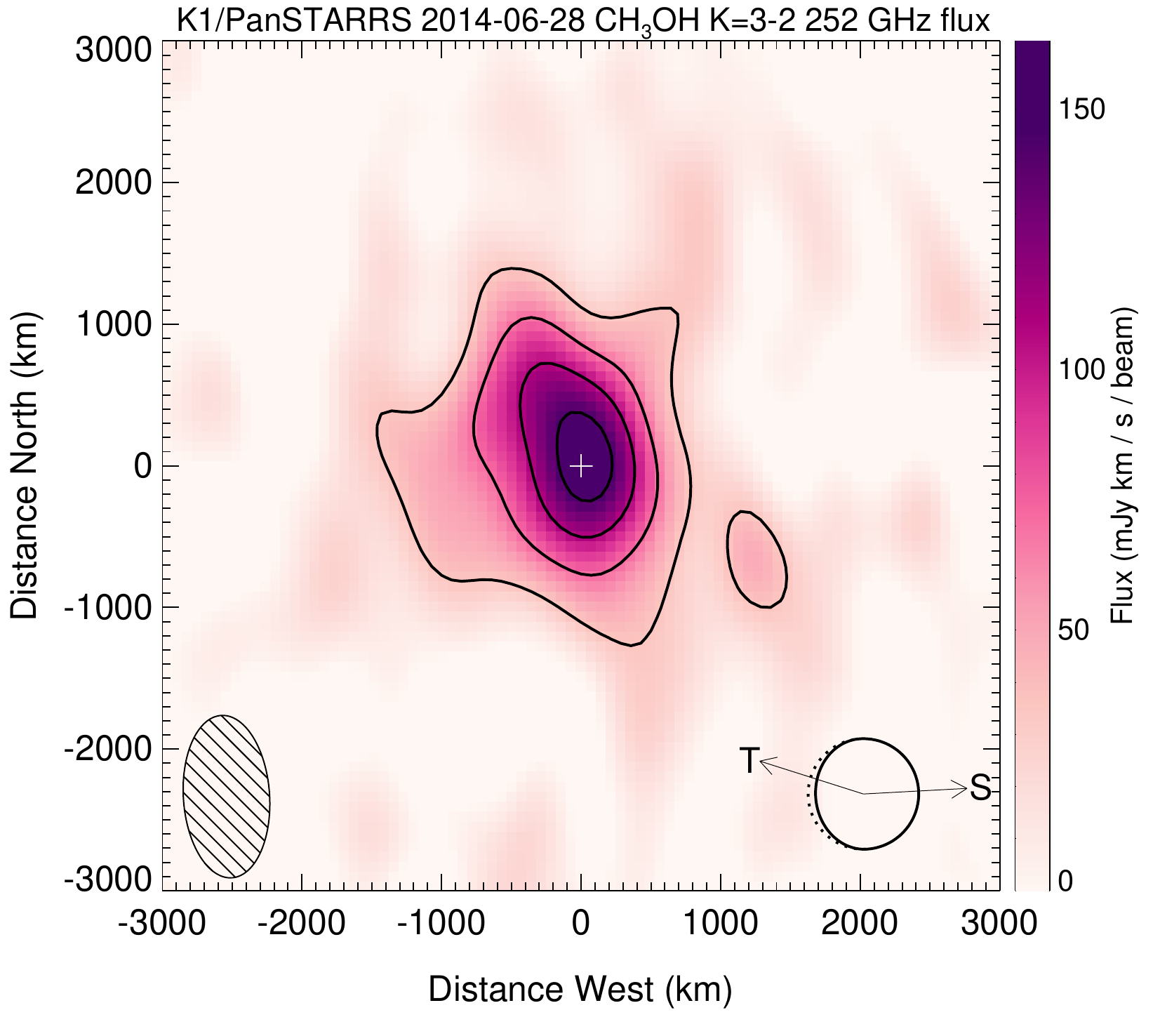}
\hspace{2mm}
\includegraphics[width=0.45\textwidth]{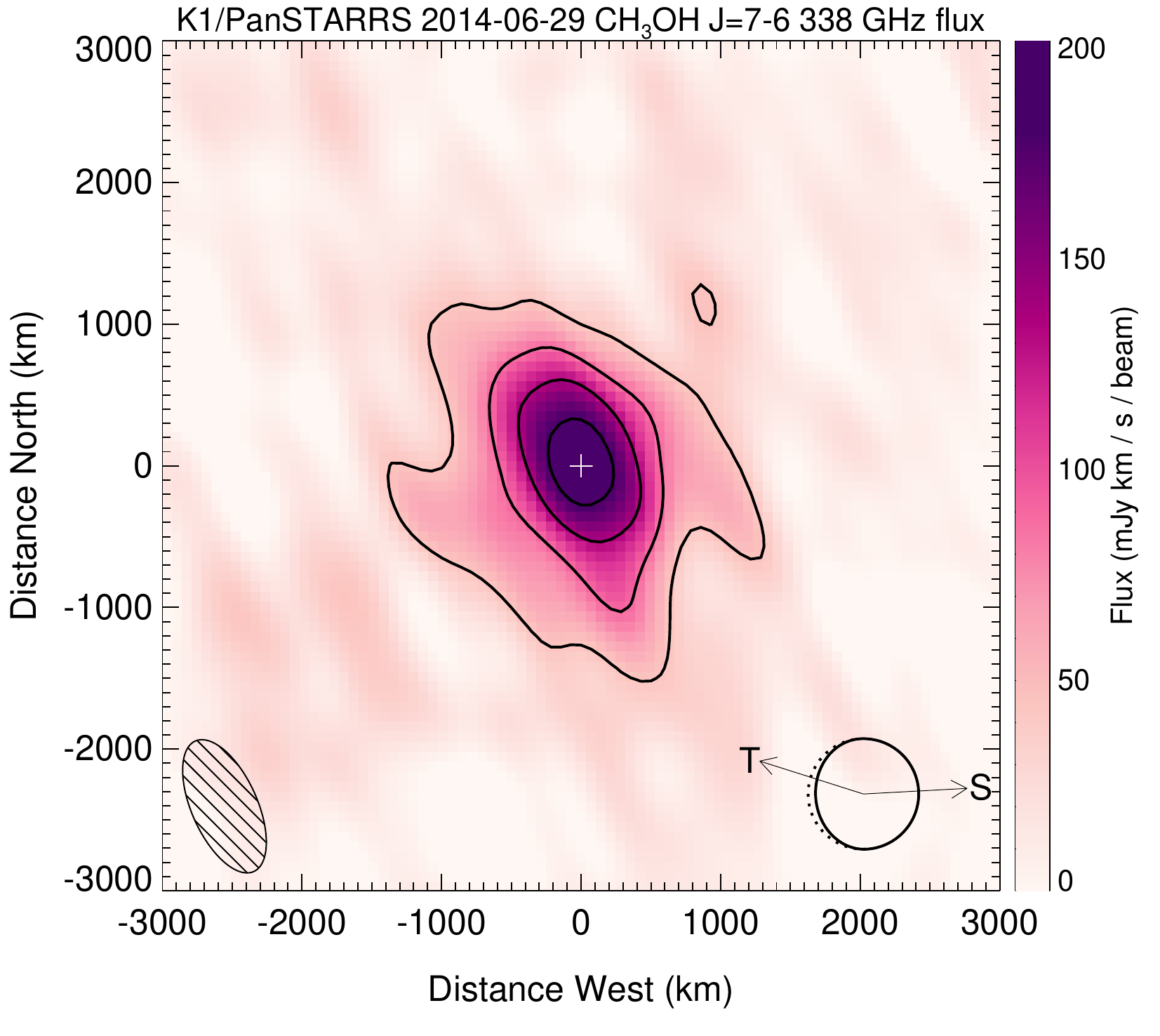}
\caption{Spectrally-integrated ALMA flux maps of CH$_3$OH in comet C/2012 K1 (PanSTARRS) observed on 2014-06-28 at 252~GHz (left) and 2014-06-29 at 338~GHz (right), integrated over all detected transitions. Coordinate axes are aligned with the RA-dec. grid, with celestial north towards the top. White crosses indicate the emission peaks,  which are employed as the origin of the coordinate axes. Contours are plotted at $3\sigma$ intervals, where $\sigma$ is the RMS noise in each map. The FWHM of the (Gaussian) restoring beams are shown lower left. Direction of the Sun (S) and orbital trail (T) with respect to the comet are indicated lower right, along with the illumination phase angle of $30^{\circ}$. \label{fig:maps}}
\end{figure*}

Forteen individual emission lines of the $K=3-2$ band were detected on 28 June and sixteen lines of the $J=7-6$ band were detected on 29 June. The complete list of detected CH$_3$OH transitions and their upper-state energies is given in Table \ref{tab:k1lines}, and the observed spectra are shown in Figure \ref{fig:spec}. These spectra were extracted by averaging the data over a circular aperture $1.5''$ in diameter, centered on the peak of the CH$_3$OH emission. Spectrally-integrated maps of the CH$_3$OH emission (shown in Figure \ref{fig:maps}) were obtained by integrating over the spectral ranges covered by the lines in Table \ref{tab:k1lines}. The CH$_3$OH emission peaks were approximately spatially coincident on both dates, offset by $1.2''$ south-west from the ALMA phase center, which is within the typical range of uncertainty for the position of the nucleus using optically-derived cometary ephemerides. No continuum emission was detected and a $3\sigma$ upper limit of 0.3~mJy was obtained for the average continuum flux (in band 7).

\begin{table}
\caption{CH$_3$OH lines detected in C/2012 K1 (PanSTARRS)\label{tab:k1lines}}
\vspace{-4mm}
\begin{center}
\begin{tabular}{lccccc}
\hline
ALMA&Transition&Freq.&$E_{u}$&$W_p$\\
Band&&(MHz)&(K)&K\,\kms\\
\hline

6& $10  _{3}      -10 _{2}\ A^{-+}    $ & 251164 & 177.5 &  0.14 (0.10)\\   
 & $9  _{3}      - 9 _{2}\ A^{-+}     $ & 251360 & 154.3 &  0.54 (0.10)\\                   
 & $8  _{3}      - 8 _{2}\ A^{-+}     $ & 251517 & 133.4 &  0.71 (0.10)\\   
 & $7  _{3}      - 7 _{2}\ A^{-+}     $ & 251642 & 114.8 &  0.62 (0.10)\\   
 & $6  _{3}      - 6 _{2}\ A^{-+}     $ & 251738 & 98.5  &  0.47 (0.10)\\   
 & $5  _{3}      - 5 _{2}\ A^{-+}     $ & 251812 & 84.6  &  1.07 (0.10)\\   
 & $4  _{3}      - 4 _{2}\ A^{-+}     $ & 251867 & 73.0  &  0.60 (0.10)\\   
 & $5  _{3}      - 5 _{2}\ A^{+-}     $ & 251891 & 84.6  &  0.78 (0.10)\\   
 & $6  _{3}      - 6 _{2}\ A^{+-}     $ & 251896 & 98.5  &  1.04 (0.10)\\   
 & $4  _{3}      - 4 _{2}\ A^{-+}     $ & 251900 & 73.0  &  0.56 (0.10)\\   
 & $3  _{3}      - 3 _{2}\ A^{-+}     $ & 251906 & 63.7  &  0.44 (0.10)\\   
 & $3  _{3}      - 3 _{2}\ A^{+-}     $ & 251917 & 63.7  &  0.38 (0.10)\\   
 & $7  _{3}      - 7 _{2}\ A^{+-}     $ & 251924 & 114.8 &  0.84 (0.10)\\   
 & $8  _{3}      - 8 _{2}\ A^{+-}     $ & 251985 & 133.4 &  0.68 (0.10)\\
\hline
7& $7  _{0 }      - 6 _{0 }\ E   $ & 338124 & 78.1     &  0.65 (0.11)\\  
 & $7  _{-1}      - 6_{-1 }\ E   $ & 338345 & 70.6     &  1.08 (0.11)\\  
 & $7  _{0 }  - 6 _{0 }\ A^+     $    & 338409 & 65.0  &  1.22 (0.11)\\
 & $7  _{-4}  - 6_{-4 }\ E       $    & 338504 & 152.9 &  0.50 (0.11)\\ 
 & $7  _{4}  - 6_{4 }\  A^-      $    & 338513 & 145.3 &  1.46 (0.11)\\ 
 & $7  _{4}  - 6_{4 }\  A^+      $    & 338513 & 145.3 &  $B$        \\ 
 & $7  _{2 }  - 6 _{2 }\ A^-     $    & 338513 & 102.7 &  $B$        \\ 
 & $7  _4     - 6 _4\ E          $    & 338530 & 161.0 &  0.32 (0.11)\\
 & $7  _{3 }  - 6 _{3 }\ A^+     $    & 338541 & 114.8 &  1.28 (0.14)\\ 
 & $7  _{3 }  - 6 _{3 }\ A^-     $    & 338543 & 114.8 &  $B$        \\ 
 & $7  _{-3}  - 6_{-3 }\ E       $    & 338560 & 127.7 &  0.39 (0.12)\\ 
 & $7  _{3 }  - 6 _{3 }\ E       $    & 338583 & 112.7 &  0.30 (0.12)\\ 
 & $7  _{1 }  - 6 _{1 }\ E       $    & 338615 & 86.1  &  0.83 (0.12)\\ 
 & $7  _{2 }  - 6 _{2 }\ A^+     $    & 338640 & 102.7 &  0.64 (0.12)\\
 & $7  _{2 }  - 6 _{2 }\ E       $    & 338722 & 87.3  &  1.63 (0.14)\\ 
 & $7  _{-2}  - 6_{-2 }\ E       $    & 338723 & 90.9  &  $B$        \\

\hline
\end{tabular}
\parbox{0.9\columnwidth}{\footnotesize 
\vspace*{1mm}
Note --- Integrated line intensities $W_p$ are given for the CH$_3$OH column density peak, with $1\sigma$ statistical errors in parentheses. $B$ denotes transitions that are blended with the line above.\\
}
\end{center}
\end{table}

The CH$_3$OH maps show a compact distribution with a strong central peak. The flux falls rapidly with distance from the center, which is consistent with the combined effects of (isotropic) outflow expansion and photodissociation due to Solar irradiation. Due to the lack of ALMA baselines shorter than 20~m, the largest angular scale detectable in our observations is $\approx7.4''$ (or $\approx10^4$~km), which results in filtering out of coma structures larger than this size. Based on a simple Haser parent model for K1/PanSTARRS (see Section \ref{sec:radial}), spatial filtering is expected to have a negligible impact on the peak CH$_3$OH flux (less than a few percent), whereas 2000~km ($1.4''$) from the nucleus, only $\sim75$\% of the predicted flux per beam may be recovered.

\subsection{Rotational diagrams}

\begin{figure*}
\centering
\includegraphics[width=0.45\textwidth]{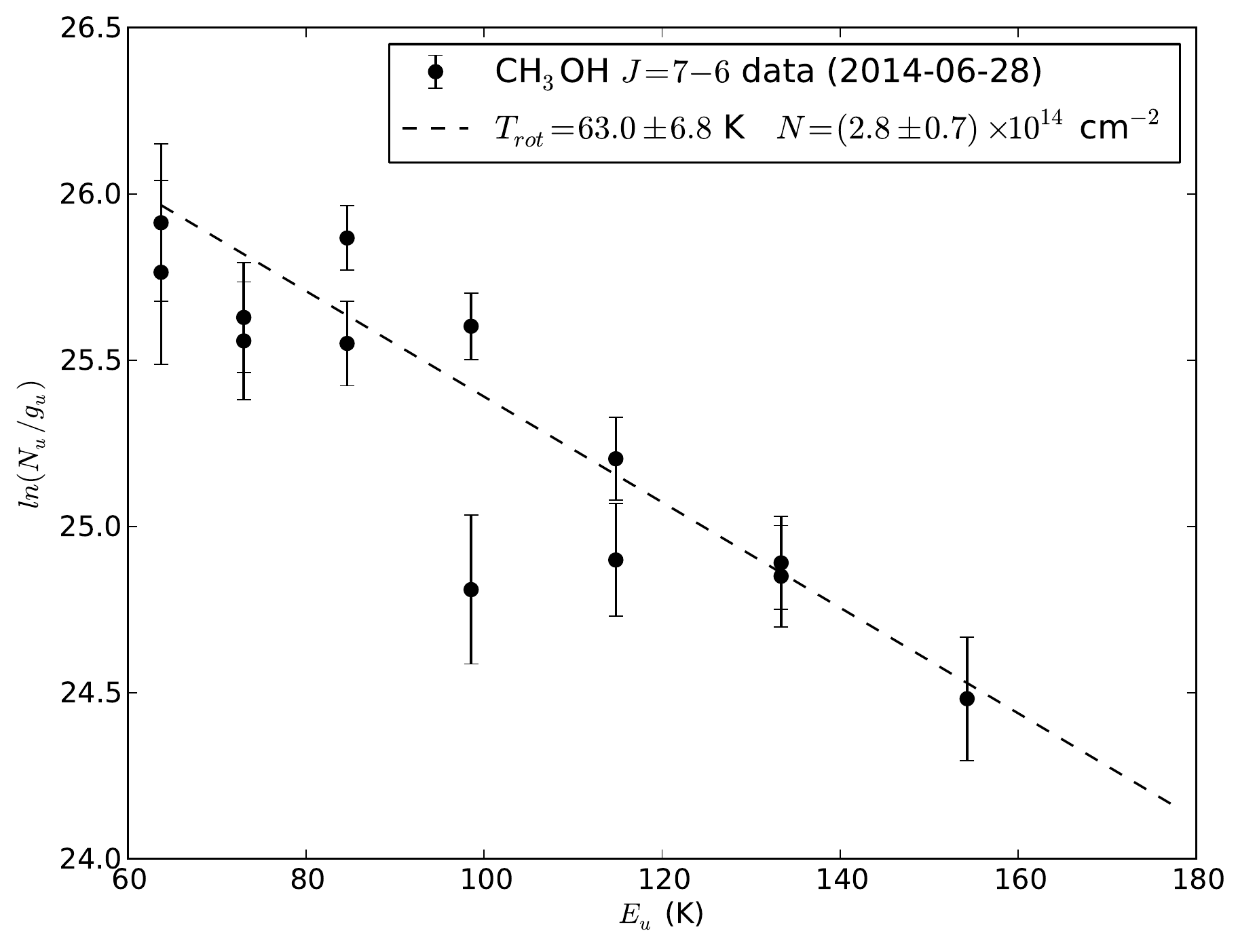}
\hspace{4mm}
\includegraphics[width=0.45\textwidth]{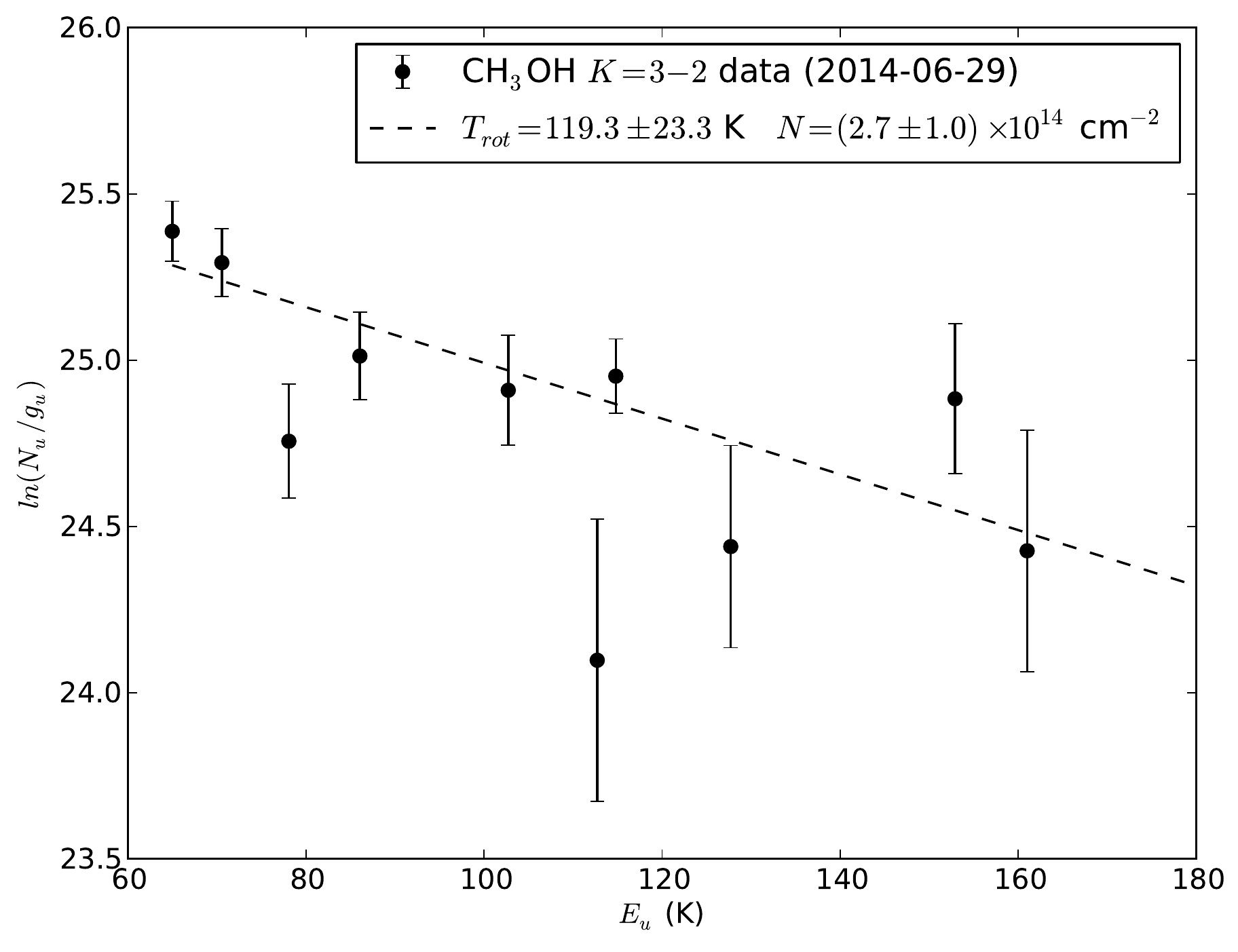}
\includegraphics[width=0.45\textwidth]{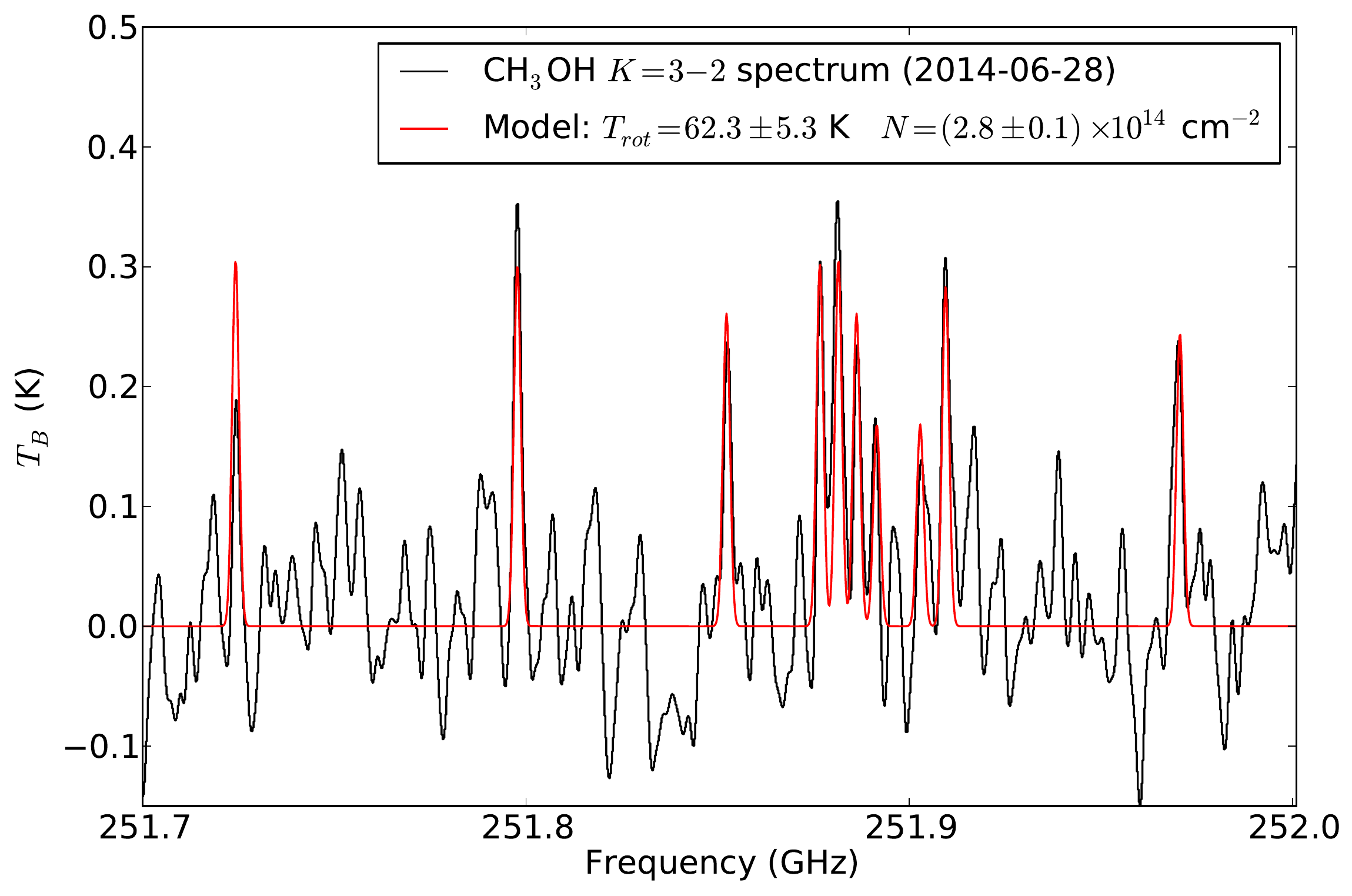}
\hspace{4mm}
\includegraphics[width=0.45\textwidth]{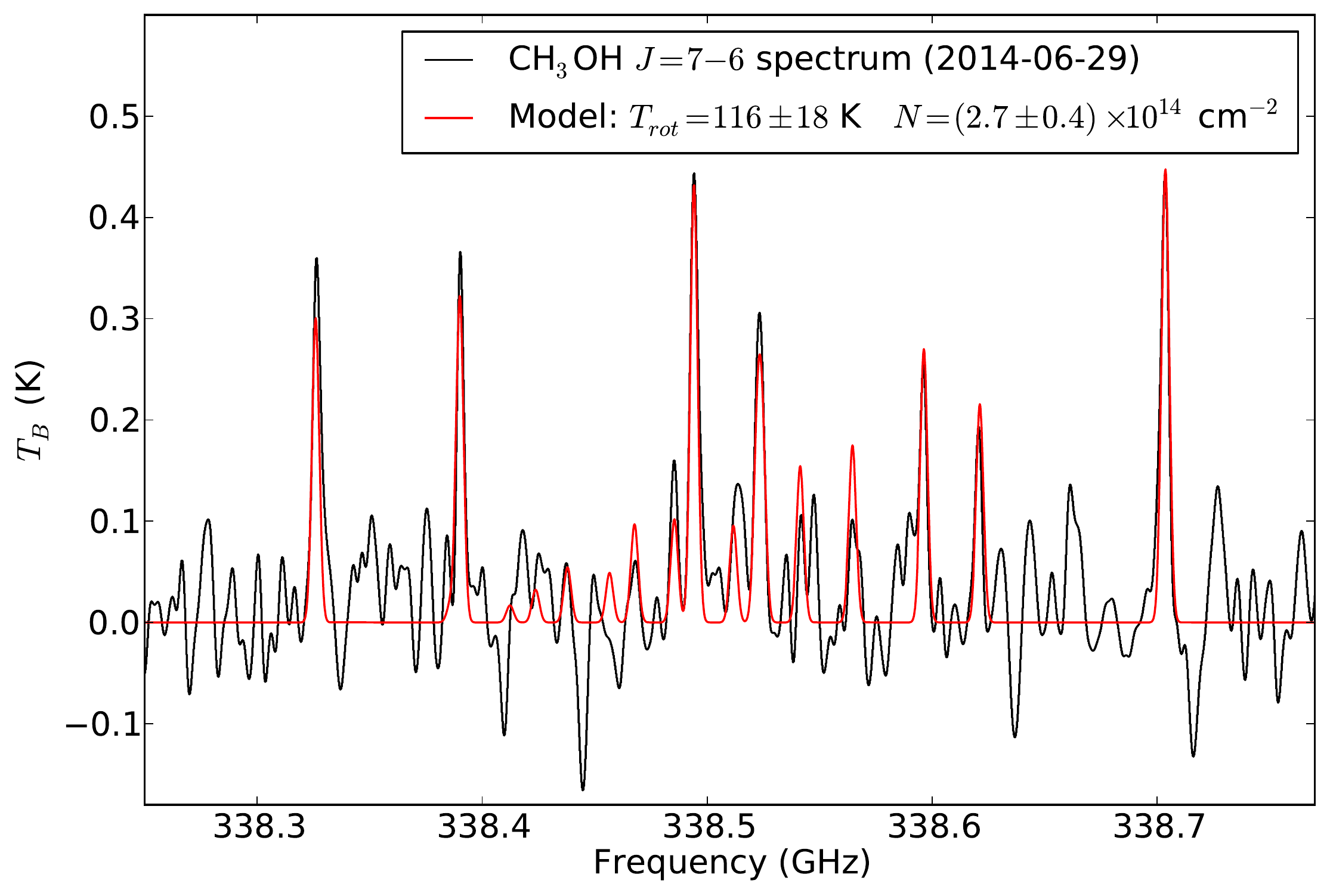}
\caption{Top: Rotational diagrams for K1/PanSTARRS CH$_3$OH, constructed using spectra obtained at the column density peaks for the $K=3-2$ transitions (left) and the $J=7-6$ transitions (right). Bottom: Least-squares model fits to the observed (smoothed) spectra; for clarity, only a portion of the entire spectral region is shown for each date.\label{fig:rotdiag}}
\end{figure*}

Detailed spectral modeling is required to interpret the observed multi-line CH$_3$OH data. We begin by using the method outlined by \citet{boc94} to construct rotational excitation diagrams to obtain the internal (rotational) temperature of the CH$_3$OH molecules ($T_{rot}$), averaged along the line of sight.  Figure \ref{fig:rotdiag} shows rotational diagrams for June 28 (left) and June 29 (right), obtained from spectra extracted at the CH$_3$OH column density peak. The integrated line intensities (along with their statistical $1\sigma$ errors) for this position are given in Table \ref{tab:k1lines}. The gradient of the rotational diagram is equal to $-1/T_{rot}$ and the intercept is equal to $ln(N/Q)$ where $Q$ is the partition function and $N$ is the column density. Errors on $T_{rot}$ and $N$ were obtained from the errors on the gradient and intercept, resulting in values of $T_{rot}=63.0\pm6.8$~K, $N=(2.8\pm0.7)\times10^{14}$~cm$^{-2}$ on June 28 and $T_{rot}=119\pm23$~K, $N=(2.7\pm1.0)\times10^{14}$~cm$^{-2}$ on June 29. No significant differences were detected between the abundances of the $A$ and $E$ nuclear spin states of CH$_3$OH, consistent with an equilibrium (high-temperature) spin distribution as observed in comet Hale-Bopp by \citet{par07}. Thus, we assumed equal abundances of $A$ and $E$ CH$_3$OH in our analysis from here on.

Unfortunately, the rotational diagram method suffers from several shortcomings:  blended lines with differing {upper-state energies} ($E_u$), and lines with signal-to-noise $<1$ cannot be included in the diagram, which limits the accuracy of the results. Furthermore, the full frequency extent of each line is not known a-priori, which hinders the accuracy of the individual line measurements --- in determining the integration widths for weak lines, a balance has to be found between including the full line flux and excluding noisy data in the line wings, which inevitably results in some reduction in the overall signal-to-noise for each line (for our data, we used integration limits of $\pm1.5$~\kms\ for all lines). In order to obtain the most reliable temperatures and column densities, spectral line modeling is employed to maximise the available information in our ALMA data.

\subsection{Line modeling}
\label{sec:modeling}

Using a modified version of the spectral line modeling algorithm previously applied to interstellar rotational emission lines by \citet{cor13}, the spectra at each point in our CH$_3$OH maps were extracted and fitted (using a nonlinear least-squares algorithm) to determine $T_{rot}$ and $N$ as a function of the spatial coordinate. The technique works by fitting Gaussian optical depth curves to each emission line (as a function of frequency), parameterised by $T_{rot}$, $N$ and the line FWHM. The observed CH$_3$OH line intensity ratios encode the rotational temperature, and the absolute intensity ($T_B$) scaling encodes the column density, so the best fit to each spectrum can be found for a unique pair of [$T_{rot}$, $N$] values, as determined by the minimum of the sum-of-squares of the residuals between the observation and model (or $\chi^2$ value). Error estimates were obtained using a Monte Carlo noise-resampling technique, whereby for each spectrum, 300 synthetic, Gaussian, random noise spectra were generated (with standard deviation equal to the RMS noise of the observed spectrum), which were then added one at a time to the (noise-free) best-fitting model spectrum. The same least-squares fitting procedure was then repeated for each of the 300 synthetic datasets. The $1\sigma$ errors on $T_{rot}$, $N$ were obtained from the $\pm68$\% percentiles of the resulting set of fit parameters.

This procedure assumes a Gaussian shape for the emission lines, which can be a poor approximation for cometary lines observed at high spectral resolution --- indeed, many of our observed CH$_3$OH line profiles show slight asymmetries (consistent with asymmetric outgassing). To avoid problems with the fitting due to such non-Gaussianity, each extracted spectrum was first convolved with a Gaussian broadening kernel of FWHM 3~\kms. This is sufficiently broad with respect to the FWHM of the observed lines ($\approx0.9-1.2$~\kms) that it effectively results in the smoothing out of their profiles, producing a line shape practically indistinguishable from Gaussian, and independent of the specific outflow geometry of the comet. This smoothing also has the benefit that any small variations in the line FWHM and Doppler shift of the gas over the ALMA field of view ($\lesssim\pm0.25$~\kms) can be neglected, allowing these values to be held fixed during the fitting to further improve the accuracy of the results, which is particularly useful for the noisier data towards the edge of the field of view.

Examples of the line modeling results are shown in the lower panels of Figure \ref{fig:rotdiag}, obtained using spectra extracted at the CH$_3$OH column density peak position. The quality of these fits is {very good} (with reduced $\chi^2$ values in the range 1.0-1.3), and the $T_{rot}$ and $N$ values derived using this method ($T_{rot}=62.3\pm5.3$~K, $N=(2.8\pm0.1)\times10^{14}$~cm$^{-2}$ on June 28 and $T_{rot}=116\pm18$~K, $N=(2.7\pm0.4)\times10^{14}$~cm$^{-2}$ on June 29) are in good agreement with those obtained using the rotational diagram method. Due to improved utilization of the noisier data, this line modeling technique results in significantly improved accuracy of the derived parameters, and is therefore adopted as the preferred method of analysis from here on. The uncertainties derived from Monte Carlo noise replication are also expected to be more reliable because they implicitly account for correlations between the errors on $T_{rot}$ and $N$, whereas in the rotational diagram analysis, the (partially correlated) errors on the gradient and intercept cannot be easily disentangled.

\subsection{Temperature and column density maps}
\label{sec:maps}

\begin{figure*}
\centering
\includegraphics[width=0.45\textwidth]{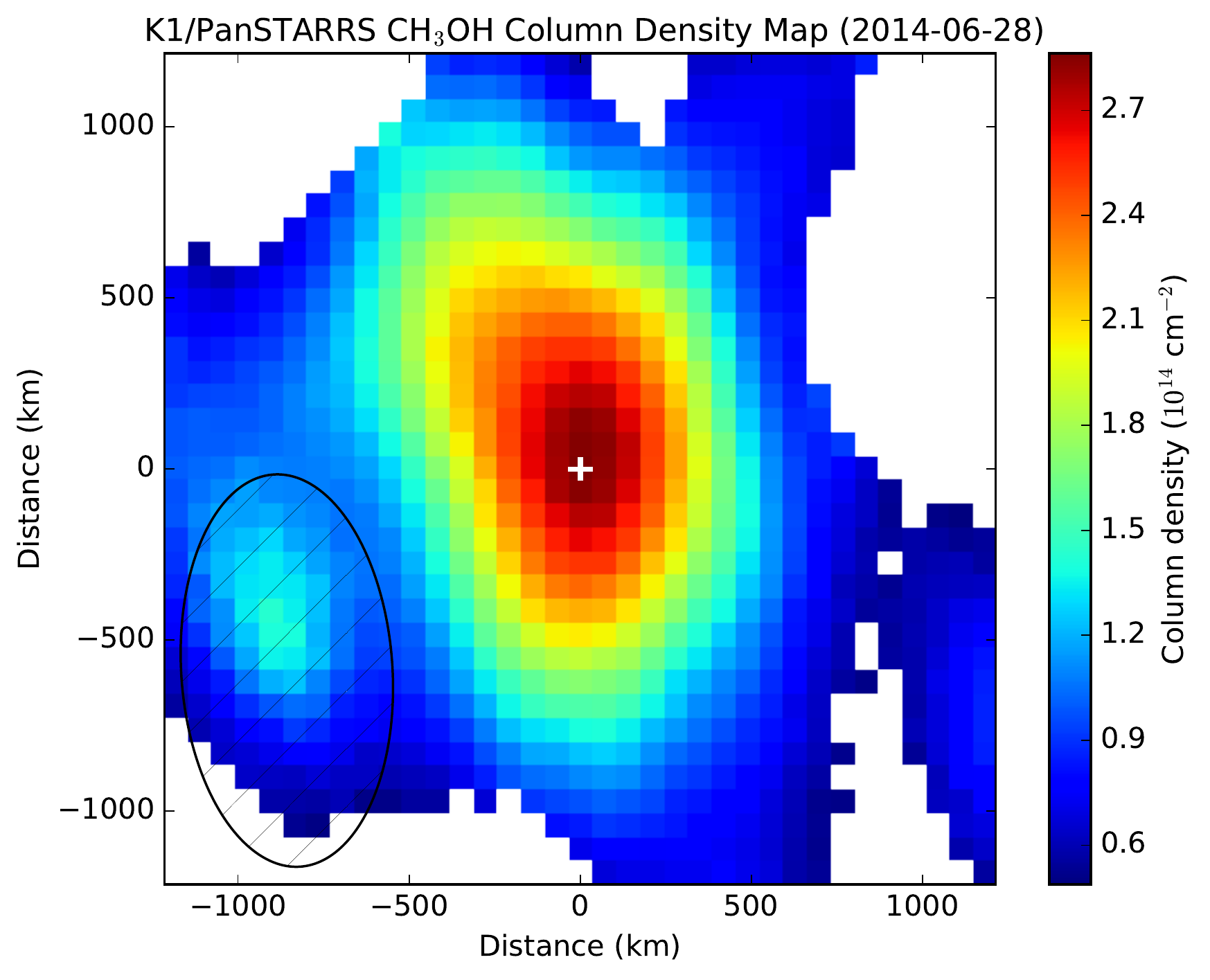}
\hspace{4mm}
\includegraphics[width=0.45\textwidth]{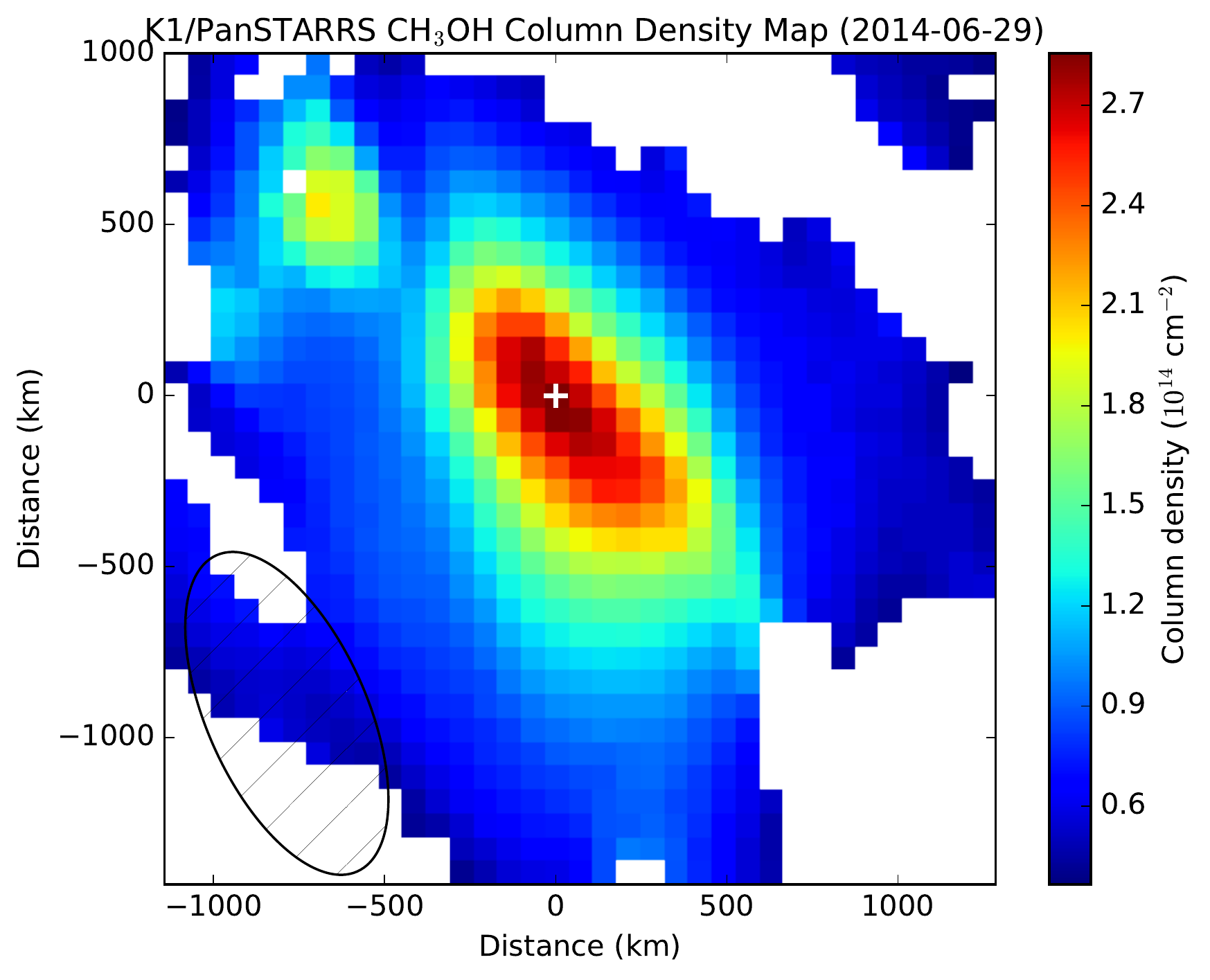}
\includegraphics[width=0.45\textwidth]{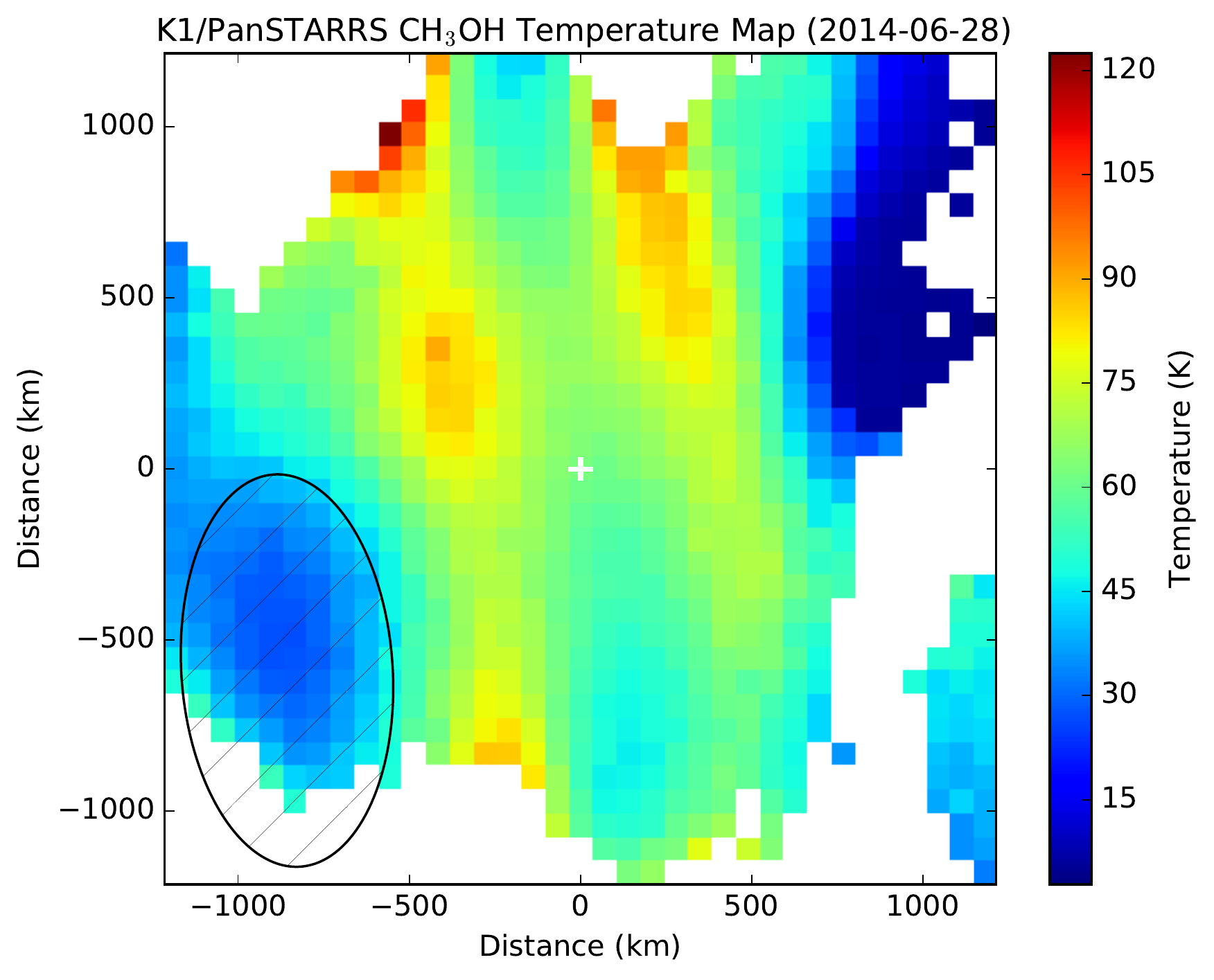}
\hspace{4mm}
\includegraphics[width=0.45\textwidth]{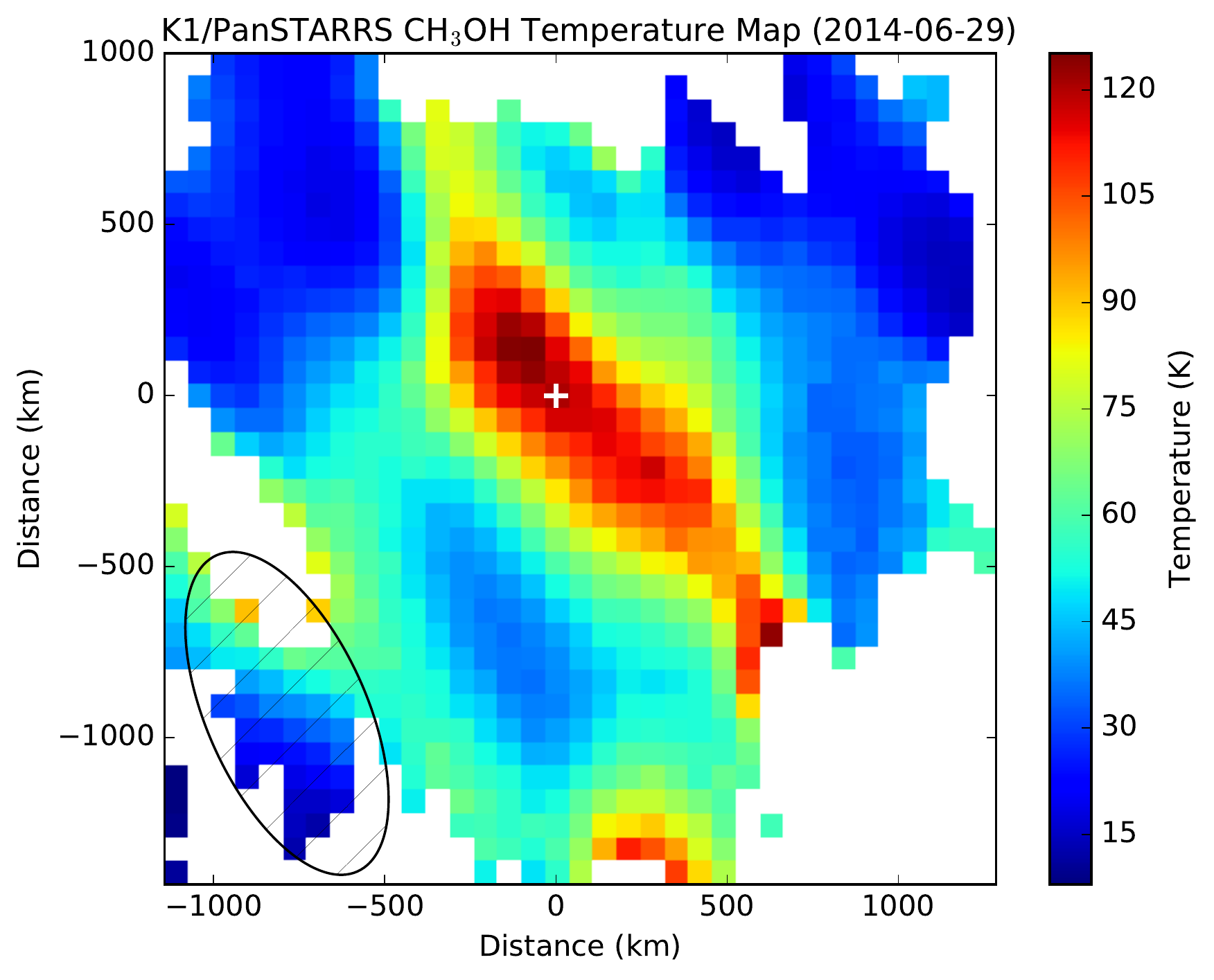}
\caption{Column density maps (top) and rotational temperature maps (bottom) for K1/PanSTARRS CH$_3$OH, derived from pixel-by-pixel spectral line modeling. Coordinate axes are aligned as in Figure \ref{fig:maps}, with celestial north towards the top. Left panels show the results for the $K=3-2$ transitions and right panels are for the $J=7-6$ transitions. Values with an uncertainty of greater than 50\% have been masked (shown as white pixels). White crosses (`+') mark the column density peak and the hatched ellipses show the instrumental resolution. The $0.05''$ pixel scale corresponds to 71~km at the distance of the comet. \label{fig:ltemaps}}
\end{figure*}

Spectra were extracted pixel-by-pixel from the CH$_3$OH data cubes within an area of $2''\times2''$ {($40\times40$ pixels)} centered on the integrated emission peak. Using the line fitting method described in Section \ref{sec:modeling}, temperatures and column densities were obtained for each pixel to create the maps shown in Figure \ref{fig:ltemaps}. These maps only show the derived $T_{rot}$ and $N$ values with uncertainties of less than 50\% --- values with larger errors have been masked (and are shown in white). The mean errors on these $T_{rot}$ and $N$ maps are $\pm14$~K and $\pm1.8\times10^{13}$~cm$^{-2}$ for June 28, $\pm15$~K and $\pm1.7\times10^{13}$~cm$^{-2}$ for June 29. Naturally, a higher CH$_3$OH column density leads to a stronger spectrum and hence lower uncertainties on $T_{rot}$ and $N$, so the regions closer to the centers of the maps are more reliable. The maps have been positioned {with their origins} at the respective column density peaks, which are marked with white crosses.

The CH$_3$OH column density maps show a single, dominant peak as expected for spherically-symmetric outgassing from a compact nucleus. The shapes of the column density peaks are strongly influenced by the elliptical telescope beam, which hinders interpretation of the true structure of the innermost coma. {On both dates, the column density maps show a significant extension towards the top left, which is in addition to the dominant outflow structure. The direction of this extension coincides approximately with the direction of the comet's orbital trail, perhaps implying some CH$_3$OH release from trailing material, although the presence of a directional CH$_3$OH jet/vent cannot be ruled out.}

Interestingly, whereas on June 29 the $T_{rot}$ distribution shows a strong central peak {about the nominal position of the nucleus} (falling from $116\pm18$~K to $\approx50\pm10$~K within about 500~km), on June 28 the main source of CH$_3$OH appears to be located within a temperature trough, which forms part of a much broader region of high temperatures that extends down the whole map in a roughly north-south direction. The central temperature on June 28 was $62\pm5$~K and the two flanking temperature maxima have $T_{rot}\approx85\pm14$~K on the left and $\approx73\pm12$~K on the right. However, the statistical significance of these peaks relative to the central temperature is low ($\sim1$-$2\sigma$), so a monotonically decreasing (or flat) temperature as a function distance from the nucleus in this region cannot be ruled out. Similarly, although some of the temperature structure towards the bottom left of the central peak on June 29 could be real, the uncertainties preclude a robust interpretation of the other features in these maps.

\subsection{Radial profiles}
\label{sec:radial}

\begin{figure*}
\centering
\includegraphics[width=0.45\textwidth]{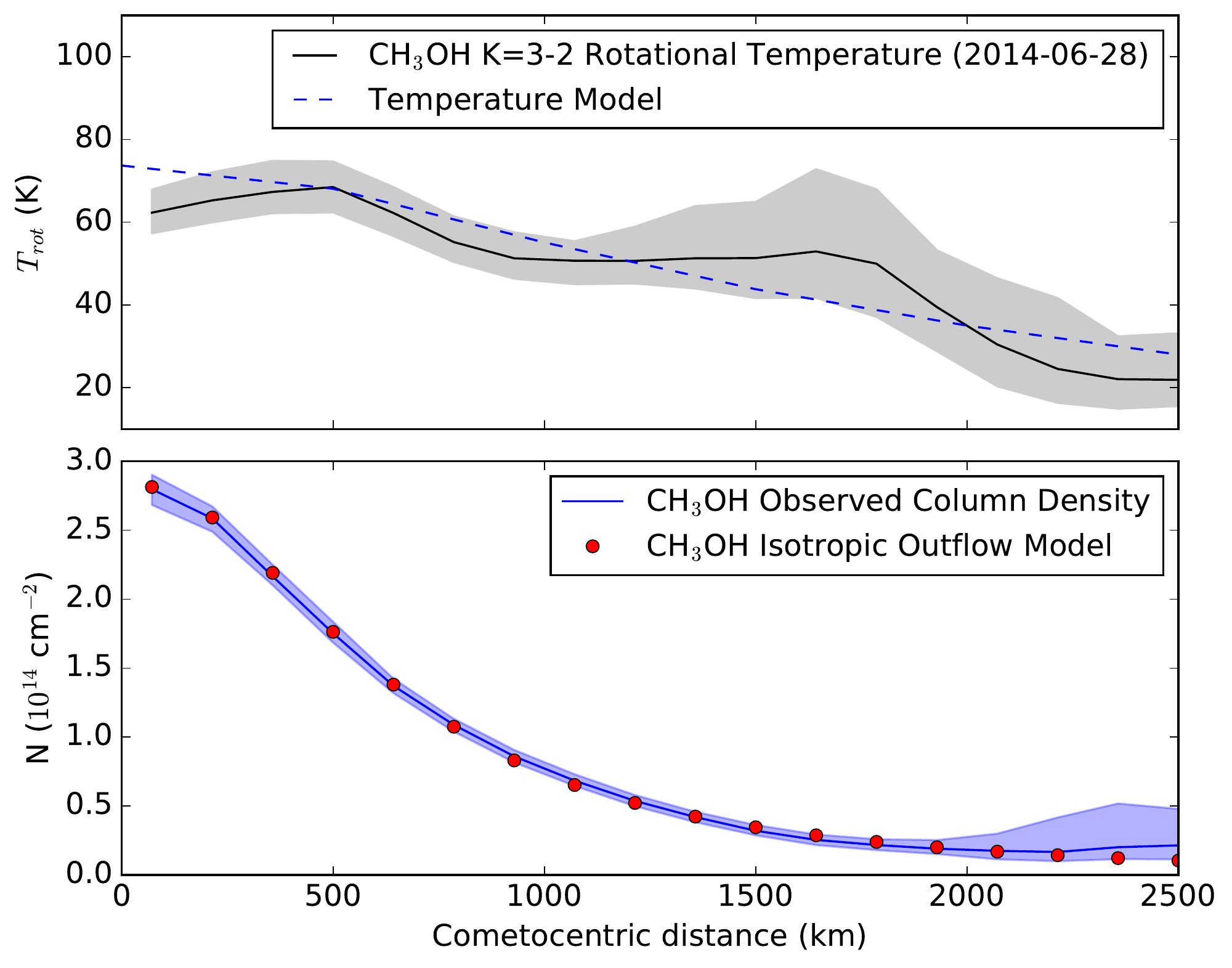}
\hspace{4mm}
\includegraphics[width=0.45\textwidth]{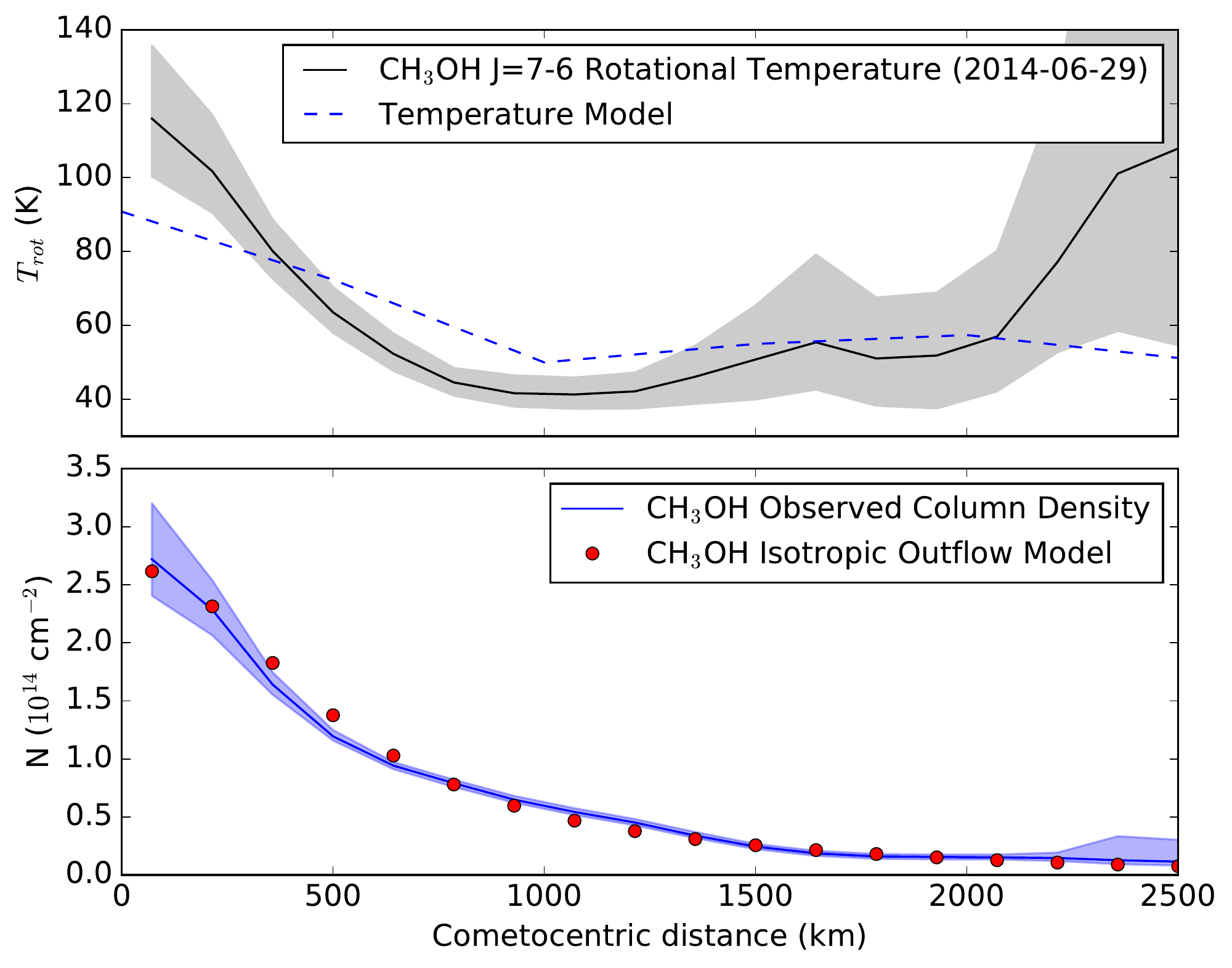}
\caption{Azimuthally-averaged CH$_3$OH rotational temperatures (upper panels) and column densities (lower panels), taken about the column density peaks labeled in Figure \ref{fig:ltemaps}. The $K=3-2$ transitions are shown (left) and the $J=7-6$ transitions (right). Simulated (Haser-type) isotropic outflow model column density profiles are overlaid with red points, corresponding to CH$_3$OH production rates of $1.9\times10^{27}$~s$^{-1}$ on June 28 and $1.4\times10^{27}$~s$^{-1}$ on June 29. {Dashed lines in the upper panels indicate the rotational temperature profiles obtained from radiative transfer modeling, using the kinetic temperature distributions shown in Figure \ref{fig:trotModel} (solid black lines)}. \label{fig:profiles}}
\end{figure*}

To help average out any clumpiness or small-scale coma structures and provide a clearer picture of the dominant physical and chemical processes as a function of distance from the nucleus, we perform azimuthal averaging of the ALMA data. Such azimuthally averaged data also benefit from a significantly improved signal-to-noise ratio, particularly at large radii, allowing us to probe the conditions further from the nucleus than in the 2-D maps. 

Taking the CH$_3$OH column density peak as the center, our data were binned into a series of $0.1''$-wide annuli, and the average flux per spectral channel was taken in each annulus, resulting in azimuthally-averaged spectra ($\bar{S_{\nu}}$) as a function of sky-projected radius $\rho$. These $\bar{S_{\nu}}(\rho)$ were subject to the same fitting procedure outlined in Section \ref{sec:modeling} to derive radial [$T_{rot}$, $N$] profiles, shown in Figure \ref{fig:profiles}.

The 1-D radial temperature profiles show a general trend for falling $T_{rot}$ as a function of distance. Similar to the 2-D maps, on June 28 $T_{rot}$ shows a possible ($1\sigma$ confidence) increase from $62\pm5$~K to $68\pm6$~K between $\rho=0$-500~km, followed by a relatively slow, steady decrease to $22\pm9$~K at $\rho=2500$~km. Conversely, on June 29 the temperature decreased sharply from $116\pm18$~K to $42\pm4$~K between $\rho=0$-1000~km. In addition to starting at least 30~K hotter near the nucleus, overall, the CH$_3$OH rotational temperatures remained at least 20~K higher throughout the observed coma on the 29th than the 28th, implying the presence of a substantial additional coma heat source on the 29th (22.5 hr later).

The possible temperature increase at distances $\rho>1500$~km on June 29 identified by \citet{cor16} is less clear in our new results; the large error bars on our $T_{rot}$ profile at $\rho>2000$~km indicate that this result may have been at least partly due to an underestimate of the errors by \citet{cor16}, who used a rotational diagram method rather than spectral line fitting. Furthermore, the azimuthal profiles in the present study were taken about the CH$_3$OH column density peak, whereas those of \citet{cor16} were taken about the integrated emission line peak; these positions were coincident on June 28, but on June 29 the column density peak is offset north-east by $0.16''$, leading to differences in the radial temperature profiles. {The elliptical shape of the telescope beam introduces some additional uncertainty into our radial temperature and column density profiles, which would be particularly problematic for any asymmetric coma features that happen to line up with the long axis of the beam. The (long-axis) spatial resolution HWHM of $550$~km on June 28 and $500$~km on June 29 should thus be considered as upper limits for the possible radial error margins on these azimuthally-averaged profiles}.

On both dates the azimuthally-averaged column densities decreased smoothly with distance from the peak value of about $2.8\times10^{14}$~cm$^{-2}$, with a shape closely consistent with uniform, spherically-symmetric outflow. This is confirmed by comparison of the observed $N(\rho)$ profiles in Figure \ref{fig:profiles} with profiles obtained from a Haser-type spherical outflow model, in which CH$_3$OH is assumed to flow isotropically from the nucleus at a constant velocity of 0.5~\kms\ (consistent with the average CH$_3$OH line Doppler FWHM of $\approx1.0$~\kms), and is photodissociated at a rate of $10^{-5}$~s$^{-1}$ \citep{hue92}. To match the observational point spread functions, the model column density maps were run through the ALMA simulator in CASA with array configuration, imaging and cleaning parameters matching those of our observationions. Good fits to the observed $N(\rho)$ profiles were obtained for CH$_3$OH production rates of $1.9\times10^{27}$~s$^{-1}$ on June 28 and $1.4\times10^{27}$~s$^{-1}$ on June 29. Using the H$_2$O production rate of $(1.05\pm0.20)\times10^{29}$~s$^{-1}$ (section \ref{sec:oh}), these results imply that CH$_3$OH was sublimated directly from the nucleus of comet K1/PanSTARRS with mixing ratios of 1.1-2.2\%. 

Our model column density profiles (red dots overlaid on the blue curves in Figure \ref{fig:profiles}) provide an excellent fit to the observations on June 28, and a reasonably good fit on June 29. As with the study of \citet{cor14}, this confirms that the Haser model can be usefully applied to azimuthally-averaged ALMA observations of the inner coma.  Slight discrepancies may be explained by the presence of clumpy structure, jets or variations in the outflow velocity. Some acceleration of the coma gas within our ALMA field of view is plausible. Indeed, the derived CH$_3$OH expansion velocity of 0.5~\kms\ in our $\sim2000$~km field of interest is smaller than the value of 0.7~\kms\ obtained for OH within a $\sim10^5$~km beam, as expected if the outflow speed increased with cometocentric distance. This possibility may be explored through detailed analysis of the CH$_3$OH outflow dynamics in a future study.

\section{Discussion}

In the range $\rho=0$-2000~km, our ALMA observations indicate, as expected, a general trend for decreasing CH$_3$OH rotational temperatures as a function of cometocentric distance in the coma.  To interpret this result requires an understanding of the meaning of $T_{rot}$. As explained by \citet{boc94,boc04}, the temperature $T_{rot}$ obtained from the analysis of CH$_3$OH rotational line ratios is representative of the distribution of internal rotational level populations. Unless the molecules are in local thermodynamic equilibrium (LTE, which occurs in the dense, inner coma through collisional equipartition), $T_{rot}$ can deviate significantly from the kinetic temperature of the gas ($T_{kin}$). The relationship between $T_{kin}$ and $T_{rot}$ depends on the competing influences of microscopic collisional and radiative processes. As a result, the prevalence of LTE depends primarily on the coma density, the spontaneous radiative decay rate of the gas and the flux of external (Solar) radiation. Within a few hundred kilometers of the nucleus where densities are high, the collisional rates between molecules are usually sufficient to maintain LTE, but as the molecules flow out into the less dense, outer coma regions (at $r_c\gtrsim1000$~km), collisions become less frequent, allowing the rotational levels to radiatively cool so that $T_{rot}$ falls below $T_{kin}$. Pumping of molecular ro-vibrational levels by Solar radiation can also have an important effect on $T_{rot}$ at large distances (beyond those considered in the present study).

\begin{figure*}
\centering
\includegraphics[width=0.45\textwidth]{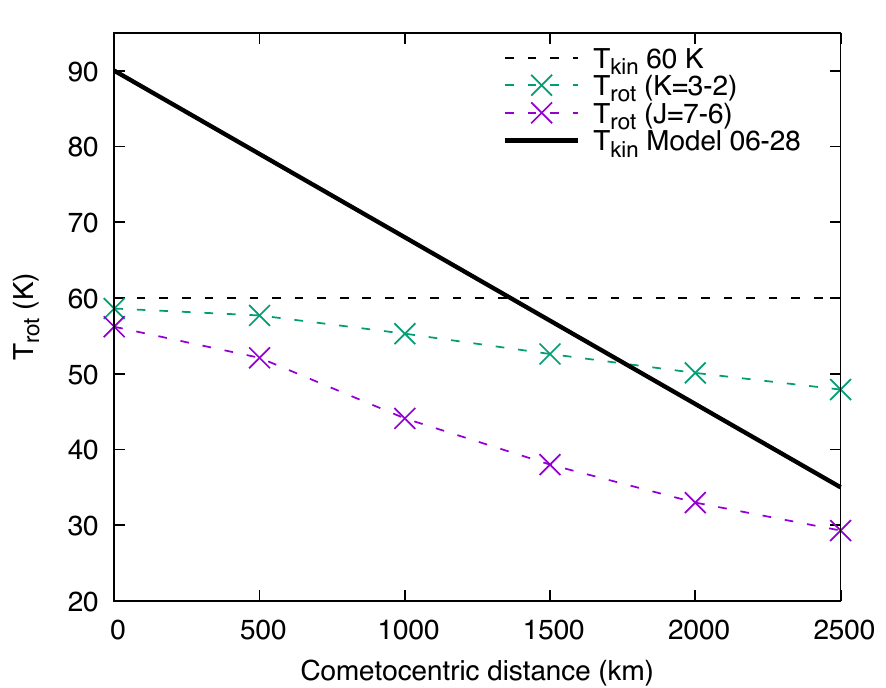}
\includegraphics[width=0.45\textwidth]{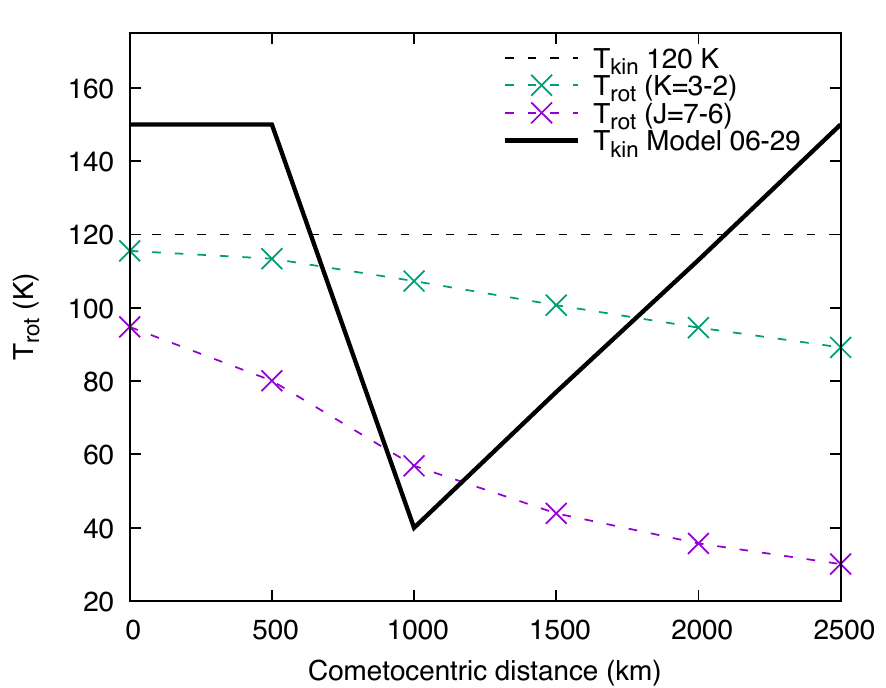}
\caption{Dashed green and purple lines show simulated CH$_3$OH rotational temperatures as a function of distance from the nucleus (using the excitation model of \citet{biv15}), for the $K=3-2$ and $J=7-6$ transitions respectively, assuming a constant coma kinetic temperature of $T_{kin}=60$~K (left panel) and 120~K (right panel). Whereas the $K=3-2$ levels remain quite close to the kinetic temperature, the $J=7-6$ levels quickly become very sub-thermal due to faster radiative cooling. {The solid black lines show our best-fitting $T_{kin}(r)$ profiles on June 28 (left) and June 29 (right).}  \label{fig:trotModel}}
\end{figure*}

Using the radiative transfer model of \citet{biv99,biv15}, which computes the CH$_3$OH level populations that result from collisions with H$_2$O and electrons, and pumping by solar radiation, we calculated the departure of $T_{rot}$ from $T_{kin}$ with distance from the nucleus. Using dashed line styles, Figure \ref{fig:trotModel} shows $T_{rot}$ as a function of projected distance ($\rho$) for the $J=7-6$ and $K=3-2$ bands, under the assumption of constant kinetic temperatures of 60~K (left panel) and 120~K (right panel). The H$_2$O production rate was taken to be $10^{29}$~s$^{-1}$ and outflow velocity 0.5~\kms, with a collisional cross section between H$_2$O and CH$_3$OH of $5\times10^{-14}$~cm$^{-2}$. In this figure, $T_{rot}(K=3-2)$ and $T_{rot}(J=7-6)$ both show increasing departures from LTE with increasing distance from the nucleus. However, whereas the $K=3-2$ band remains relatively close to LTE (with $T_{rot}>0.75T_{kin}$ for $\rho<2500$~km), for the $J=7-6$ band, $T_{rot}$ falls much rapidly with distance, reaching $0.25T_{kin}$ by $\rho=2500$~km in the 120~K model. This is due to the larger Einstein $A$ coefficients of the $J=7-6$ transitions, which result in more rapid rotational cooling. Although this trend is qualitatively similar to the observed $T_{rot}$ pattern for the two CH$_3$OH bands in Figure \ref{fig:profiles}, the observed $T_{rot}$ curves both fall significantly more rapidly than expected with a coma kinetic temperature that remains constant as a function of distance. Sub-thermal excitation is therefore insufficient to fully explain the observed $T_{rot}$ behaviour in comet K1/PanSTARRS, {and a variable $T_{kin}(r)$ profile is required.} 

{A variety of different kinetic temperature profiles were tested in our radiative transfer model, with the aim of reproducing the general behaviour of the observed $T_{rot}(r)$ profiles. For the $K=3-2$ profile on June 28, a good fit to the observed $T_{rot}(r)$ was obtained using a single-slope $T_{kin}(r)$ profile, starting at 90~K at the nucleus and falling to 35~K at $r_c=2500$~km (shown by the solid black line in the left panel of Figure \ref{fig:trotModel}). For the $J=7-6$ data from June 29, however, it was more difficult to obtain a good fit to the measured $T_{rot}(r)$ data. Even with a rapidly falling $T_{kin}(r)$ profile in the inner coma, the very steep initial drop in $T_{rot}(r)$ could not be accurately reproduced. This is partly due to line-of-sight averaging, because cooler, more distant parts of the coma can dominate the temperature contribution from a compact, warm inner region (combined with the need to keep the inner coma temperatures physically reasonable). Our best fit to the $T_{rot}(J=7-6)$ profile was obtained using the $T_{kin}(r)$ shown in black in the right panel of Figure \ref{fig:trotModel}, beginning at a relatively high temperature of 150~K, falling to 40~K at $r_c=1000$~km, then rising back to 150~K by $r_c=2500$~km. Difficulties in modeling the $T_{rot}(r)$ profile on June 29 may be due to coma asymmetries or other physical factors not included in our model. Future attempts to more robustly derive the coma kinetic temperatures from these data may require modeling the coma structure in 3-D, as well as exploring the effects of variations in electron density profiles and temperatures, and gas collisional cross sections.}

Compared with previous generations of chemical/hydrodynamic coma models (see \citealt{rod04} for a review), the temperature trends observed in K1/PanSTARRS differ from past theoretical expectations. In those models the neutral gas temperature falls rapidly as it travels outward from the nucleus, starting at $\sim100$~K and reaching a minimum of $\sim10$~K by 100~km as a result of adiabatic cooling. Around a cometocentric distance $r_c\sim100$~km, the onset of photoionisation due to Solar irradiation results in the production of a population of fast (hot) electrons, ions and neutrals. Collisions with these hot photoproducts heats the parent gases. However, the strong heating trends apparent in these conventional multi-fluid hydrodynamic models --- which predict temperatures above a few hundred kelvins for cometocentric distances $>1000$~km --- primarily reflects the fact that the neutral gas temperature is obtained by calculating the weighted average temperature over all coma neutrals, including the fast-moving photoproducts (primarily O, H, H$_2$ and OH from the photolysis of H$_2$O). Because the kinetic temperatures of these photoproducts can reach thousands of kelvins in the outer coma, their presence creates a strong bias in the average temperature of neutrals, which is then no longer properly representative of the temperature of parent gases.  To account for this problem, the temperatures of the parents and photoproducts must be treated separately, as in the direct simulation Monte Carlo (DSMC) models of \emph{e.g.} \citet{ten08} and \citet{fou12}. These models show a clear separation in the kinetic temperatures of parents and photoproducts and confirm the expectation that parent species are dominated by adiabatic cooling for distances $\lesssim10^5$~km. 

However, standard DSMC models still predict a very rapid drop in the temperature of parent molecules with distance that is inconsistent with our ALMA CH$_3$OH observations. For example, in the model for comet 67P/Churyumov-Gerasimenko by \citet{ten08}, the H$_2$O temperature falls by an order of magnitude within 100~km of the nucleus, {whereas on June 28, $T_{kin}(r)$ in K1/PanSTARRS fell by only a few percent over this range}. To resolve this discrepancy, an additional source of coma heating is required in the models. \citet{fou12} considered the sublimation of (dirty) ice grains, which can significantly raise the H$_2$O rotational temperature in the region $r_c\lesssim1000$~km provided their mass represents a significant fraction of the total gas-phase H$_2$O production. Such sublimation heating could also raise the CH$_3$OH rotational temperature, either through injection of a source of heated CH$_3$OH into the coma, or by collisions with the heated H$_2$O molecules. Other possible heating sources that may be considered in future models include suprathermal electrons and ions that could be produced through interaction of the coma with the solar wind, UV and X-rays.

The temperature trend in K1/PanSTARRS is compatible with the results of other observational studies. The steadily decreasing $T_{kin}$ behaviour found on June 28 is comparable to the linear temperature slope deduced for the inner few thousand kilometers of comet P1/Garrad \citep{boc12}. {Further, on June 29 the falling kinetic temperature in the inner coma, rising back to $\sim150$~K at larger radii is qualitatively similar to that study}.  Our observations on both dates are also broadly consistent with the general trend for decreasing H$_2$O rotational temperatures with distance over scales $\rho\sim10-1000$~km in comets C/2004 Q2 (Machholz) and 73P-B/Schwassmann-Wachmann 3 \citep{bon07,bon08}. Comets 103P/Hartley 2 and C/2012 S1 (ISON) showed a more complex temperature behavior. At $r_H=0.53$~AU, comet ISON's $T_{rot}$ fell from $\sim120$ K to $\sim85$ K within 1000~km of the nucelus, whereas at $r_H=0.35$~AU, evidence for a double-peaked temperature structure was observed, rising to maxima at cometocentric distances $\sim500-1000$~km \citep{bon14}. For 103P/Hartley 2, $T_{rot}$ was observed to decrease with distance on the sunward side of the nucleus, whereas on the anti-sunward side (as projected on the sky plane at a phase angle of $54^{\circ}$), evidence was found for a significant increase in $T_{rot}$ between $\rho=0$-75~km \citep{bon13}. This $T_{rot}$ behavior is analogous to the possible double-peaked temperature structure identified in K1/Panstarrs on 28 June (Figure \ref{fig:ltemaps}). As discussed by \citet{bon14}, it may be that such unexpected temperature peaks in the coma can arise as a result of photolytic or sublimation heating, but additional theoretical studies will be required to confirm this hypothesis.
 
The behavior of the coma temperature as a function of time can provide more information on the nature of the coma heating processes, and is possible because of the 22.5~hr separation between our $K=3-2$ and $J=7-6$ CH$_3$OH observations. Although our excitation model shows that $T_{rot}(J=7-6)$ may still be somewhat sub-thermal in the inner coma, the $J=7-6$ and $K=3-2$ bands are both expected to be close to the coma kinetic temperature at around $r_c=0$.  The implied dramatic increase in coma kinetic temperature between June 28th and 29th is therefore surprising (given the expected relative constancy of Solar radiation input), and is likely to have been caused by an increase in the heating rate of the inner coma, for instance, due to an increase in the supply of hot electrons or sublimating icy grains. {If the electron temperature dropped below the threshold for dissociative electron impacts with H$_2$O, this could also result in a sudden increase in the coma heating rate, as a larger fraction of the electron energy can then be converted into kinetic energy \citep[as discussed by][]{bod16}. Interactions with the Solar wind could also lead to short-timescale variability in the coma energy balance}. 

Large variations in the CH$_3$OH rotational temperature were observed over timescales of several hours in comet 103P/Hartley 2 by \citet{dra12}. {These variations were explained as due, in part, to the theorized correlation between water production rate and coma heating efficiency \citep[see][]{boc87,com04}}.  However, {the relative constancy of the CH$_3$OH production rate during our observations appears to rule out this possibility in K1/PanSTARRS, and} the full explanation for such strong, transient variations in coma heating requires further investigation.

Several review studies have drawn comparison between interstellar, protostellar and cometary ice abundances (\emph{e.g.} \citealt{ehr00}; \citealt{mum11}; \citealt{boo15}). To-date, every coma species detected in the radio has also been found in interstellar clouds, and the dominant cometary ice constituents (CO, CO$_2$ and CH$_3$OH) show abundances with respect to H$_2$O that are generally within the range of values observed in protostellar environments. \citet{boc94} identified the similarity between cometary and interstellar CH$_3$OH/H$_2$O ice ratios, and our values of 1.1-2.2\% in C/2012 K1 (PanSTARRS), compared with 1-30\% in low-mass protostars \citep{mum11}, confirm this result.   Recent detailed models for disk gas and ice chemistry confirm the plausibility of a close chemical relationship between cometary and protoplanetary material --- \emph{e.g.} \citet{dro16} predict CH$_3$OH/H$_2$O~$\sim1$-4\% in the mid-plane ices for low-mass protoplanetary disks. The CH$_3$OH abundance in K1/PanSTARRS is also consistent with the detection of this molecule for the first time in a protoplanetary disk by \citet{wal16}, who obtained a gas-phase CH$_3$OH/H$_2$O ratio $\sim1$-5\% in the disk surrounding the low-mass TW Hya system (at a distance of 54~pc). The fact that CH$_3$OH appears to be depleted {in comets and protoplanetary disks compared with} the median protostellar abundance of 6\% in the nearby Galaxy \citep{boo15}, {implies that significant processing of interstellar/protostellar envelope material occurs during or after its passage to the accretion disk, thus confirming the importance of cometary ices as a record for the physical and chemical processes occurring during the formation of the Solar System.}

\clearpage
\section{Conclusion}

Using ALMA observations of C/2012 K1 (PanSTARRS), the first instantaneous spatial/spectral maps of CH$_3$OH rotational emission have been obtained in a cometary coma. Through rotational excitation analysis, 2-D spatial maps of the CH$_3$OH column density and rotational temperatures averaged along the line of sight have been derived, revealing new information on the physics and chemistry of the coma on scales 500-5000~km. We find that the $T_{rot}(J=7-6)$ and $T_{rot}(K=3-2)$ radial profiles both exhibit a relatively rapid drop with distance, which cannot be explained purely through sub-thermal excitation and must therefore be due to falling coma kinetic temperatures with distance from the nucleus. The observed temperature behavior is more consistent with the DSMC model of \citealt{fou12} (that includes coma heating from sublimating dirty ice grains), than the behavior seen in standard multi-fluid models, highlighting a need for continued research into coma heating (and cooling) mechanisms.

The CH$_3$OH radial column density profile is in good agreement with spherically-symmetric, uniform outflow from the nucleus. Accordingly, no evidence is found for significant production of CH$_3$OH in the coma, either from icy grain sublimation or photochemistry. The CH$_3$OH mixing ratios of 1.1-2.2\% in K1/PanSTARRS are consistent with previous observations of comets at infrared and radio wavelengths. Combined with the observation of CH$_3$OH outgassing directly from the nucleus, our results confirm the utility of radio interferometric observations as a probe for the abundances of complex organic molecules in cometary ice.  Our CH$_3$OH abundance adds to the evidence confirming a close chemical similarity between protostellar/protoplanetary and cometary ices.

The derivation of accurate coma {kinetic} temperatures (using CH$_3$OH or other molecules with a high density of rotational lines in the mm/sub-mm such as H$_2$CO or CH$_3$CN), combined with detailed theoretical modeling, is necessary to provide new constraints on the coma physics and further elucidate the gas heating and cooling mechanisms. Continued ALMA observations {(including observations at higher angular resolution and higher sensitivity),} will therefore play a crucial role in improving {our knowledge of coma energetics}, as well as leading to higher accuracy in cometary molecular production rates, parent scale lengths and gas outflow velocities. Improved accuracy and statistics in measurements of cometary ice abundances are key requirements for ongoing studies on the origin and evolution of icy materials in planetary systems.

\acknowledgments
This work was supported by NASA's Planetary Atmospheres and Planetary Astronomy Programs and by the National Science Foundation under Grant No. AST-1616306. It makes use of the following ALMA data: ADS/JAO.ALMA\#2013.1.01061.S. ALMA is a partnership of ESO (representing its member states), NSF (USA) and NINS (Japan), together with NRC (Canada) and NSC and ASIAA (Taiwan), in cooperation with the Republic of Chile. The Joint ALMA Observatory is operated by ESO, AUI/NRAO and NAOJ. The National Radio Astronomy Observatory is a facility of the National Science Foundation operated under cooperative agreement by Associated Universities, Inc. The Nan\c{c}ay Radio Observatory is the Unit{\'e} scientifique de Nan\c{c}ay of the Observatoire de Paris, associated as USR No. B704 to the CNRS. The Nan\c{c}ay Observatory also gratefully acknowledges the financial support of the Conseil r{\'e}gional of the R{\'e}gion Centre in France. We acknowledge the advice of Dr. Boncho Bonev on the measurement of spatial variability in coma rotational temperatures.

{\it Facilities:} \facility{Atacama Large Millimeter/submillimeter Array}

\bibliographystyle{aasjournal}
\bibliography{refs}    

\end{document}